\documentclass[aps,prb, twocolumn,floatfix,notitlepage, superscriptaddress,longbibliography]{revtex4-2}
\usepackage[utf8]{inputenc}
\usepackage{amsfonts,amsmath,amssymb,graphicx,epstopdf,verbatim}
\usepackage{mathbbol}
\usepackage{rotating}
\usepackage{epstopdf}
\usepackage{xcolor}
\usepackage{natbib}
\usepackage[colorlinks]{hyperref}
\definecolor{darkred}{rgb}{0.5,0,0}
\definecolor{darkgreen}{rgb}{0,0.5,0}
\definecolor{darkblue}{rgb}{0,0,0.5}
\hypersetup{colorlinks,
linkcolor=darkred,
filecolor=darkgreen,
urlcolor=darkblue,
citecolor=darkgreen}
\usepackage{braket}
\usepackage{graphicx}

\begin{document}

\title{Many-body quantum heat engines based on free-fermion systems}

\author{Vincenzo Roberto Arezzo}
\affiliation{SISSA, Via Bonomea 265, I-34135 Trieste, Italy}

\author{Davide Rossini}
\affiliation{Dipartimento di Fisica dell’Università di Pisa and INFN, 
Largo Pontecorvo 3, I-56127 Pisa, Italy}

\author{Giulia Piccitto}
\affiliation{Dipartimento di Matematica e Informatica, Università di Catania, Viale Andrea Doria 6, I-95128, Catania, Italy}

\begin{abstract}
  We study the performances of an imperfect quantum many-body Otto engine based on free-fermion systems.
  Starting from the thermodynamic definitions of heat and work along ideal isothermal, adiabatic, and isochoric transformations,
  we generalize these expressions in the case when the hypotheses of ideality are relaxed (i.e., nonperfect thermalization
  with the external baths, as well as nonperfect quantum adiabaticity in the unitary dynamic protocols).
  These results are used to evaluate the work and the power delivered by an imperfect quantum many-body heat engine in a finite time,
  whose working substance is constituted by a quantum Ising chain in a transverse field:
  We discuss the emerging optimal working points as functions of the various model parameters.
\end{abstract}

\maketitle

\section{Introduction}
\label{sec:intro}

Thermodynamics is one of the most fascinating topics in physics. 
Since the Industrial Revolution, the necessity of understanding mechanisms underlying the conversion of the heat in useful work
has focused efforts of the researchers in this direction, leading to the statement of the three laws of thermodynamics.
Much more recently, with the advent of quantum technologies and the miniaturization of the devices exchanging heat and work
down to the nanoscale, it has become relevant to understand these issues in the novel scenario of quantum mechanics. 
Unfortunately, while the theory of classical thermodynamics is well posed, its generalization to the quantum realm raises
some conceptual puzzles and is still an open research field.
Seminal works along this direction, proposing a generalization of the concepts of heat and work
for quantum systems~\cite{Alicki_1979, Kosloff1984}, date back to the eighties.

Only recently, the renewed interest in the possibility of quantum heat engines has led to a prolific scientific production,
starting from the pioneer proposal of the maser as a first example of quantum engine~\cite{Scovi_1959}.
A quantum heat engine is simply defined as an engine constituted by a quantum systems that can produce work
undergoing a suitable thermodynamic cycle. 
A series of papers~\cite{Quan_2007, Rossnagel_2014, hartmann2020A, Kosloff-rev2017} have proposed possible realizations
of few-body heat engines, based both on the Carnot cycle~\cite{Bender2000, bender2002entropy} and
on the Otto cycle~\cite{Solfanelli_2020,Henrich2007, Uzdin2014, Leggio2016}. Some experimental realizations have been obtained
by means of different platforms~\cite{Rossnagel_2016, TrappedIons2019, vonLindenfels2019, Peterson2019, Sheng2021, Bouton2021},
as also discussed in Ref.~\cite{Myers22AQS4}.

A promising direction to investigate quantum advantage is that of engines whose working substance is constituted
by quantum many-body systems~\cite{Mukherjee2021, Cangemi2023}. Although this research line is hindered by technical difficulties,
from both analytical and numerical point of views, some preliminary results have been proposed, both with gases
of interacting atoms~\cite{jaramillo2016quantum, Bengtsson2018, Chen2019, Fogarty2020, Carollo2020, Boubakour2023, Watson2023}
and with interacting quantum spins~\cite{Halpern2019, hartmann2020B, Wang2020, delcampo2020, Piccitto_2022, Solfanelli_2023, Williamson2024}.
It has also been shown how criticality may have an impact on the enhancement
of the engine performance~\cite{Campisi_2016, Fogarty2020, Piccitto_2022}.
Unless some exceptions~\cite{DelGrosso2022}, all results in the literature have been obtained for ideal engines
undergoing perfect thermodynamic transformations.

In this paper we study a many-body quantum Otto engine with a free-fermion medium,
in which the assumption of perfect thermodynamic transformations can be relaxed. 
We first exploit the exact integrability of the model through the Bogoliubov de-Gennes formalism
to derive some general analytic expressions for the heat exchanged and the work produced by a free-fermion system
of arbitrary size along some paradigmatic thermodynamic processes:
the nonperfect isochoric (static Hamiltonian system, coupled to an external bath)
and the nonperfect adiabatic (isolated quantum system) transformation separately,
as well as the nonperfect adiabatic transformation followed by a nonperfect isochoric one.
To model nonunitary processes, we describe the microscopics of the system-bath dynamics through a thermal bath of
harmonic oscillators at a given temperature, quadratically coupled to the system, by means of
a nonlocal Lindblad master equation ensuring thermalization at long times.

In the second part of the paper, we apply the above analytic results to the quantum Otto cycle, with the aim of finding
the best parameters which optimize the engine performances. In particular, we numerically analyze the work output and the power
delivered by the Ising quantum Otto engine when quantum adiabaticity is violated, thus generalizing
the results obtained by two of us in Ref.~\cite{Piccitto_2022} for the ideal scenario and at finite size.
We observe the emergence of a nonmonotonic behavior in the power with the various parameters of our model,
thus signalling the existence of an optimal working point.

The paper is structured as follows.
In Sec.~\ref{sec:ideal_heat-and-work}, we recap the basic concepts of quantum thermodynamics and clarify the differences
between classical and quantum realms. In Sec.~\ref{sec:td_fermionic_systems}, we derive the expressions for the heat and the work
along ideal isotherm, adiabatic, and isochoric transformations, for a generic free-fermion system that can be coupled
to a thermal bath.
We then introduce the quantum Otto cycle (Sec.~\ref{sec:The quantum Otto cycle}), whose working substance
is a free-fermion system, and analyze the performance of such engine
operating under the assumption of nonperfect thermalization and/or quantum adiabaticity.
In Sec.~\ref{sec:result_ising}, we present our numerical simulations for the case of an Ising Otto engine when
nonadiabatic transformations are taken into account. Finally, in Sec.~\ref{sec:concl} we draw our conclusions.
The appendixes provide further details on the system-bath coupling scheme we adopt within the Lindblad
formalism (App.~\ref{app:thermalization}), the ideal Ising Otto engine in the thermodynamic limit (App.~\ref{App:ising_tdlimit}),
and the way in which the Hamiltonian control parameter should be varied to implement a proper Otto cycle
producing useful work (App.~\ref{app:parametri_ising}).

\section{Thermodynamics of quantum systems}
\label{sec:ideal_heat-and-work}

One of the basic principles governing the exchange of heat and work between
two or more systems is the first law of thermodynamics,
\begin{equation}
  dE = \delta Q - \delta W, 
  \label{eq:1st_law}
\end{equation}
which formally assesses the energy conservation principle:
the variation of internal energy $E$ of a given system can be interpreted as the difference between
the heat exchanged with the external environment and the work performed by the system itself.
In the following, we adopt the convention that $\delta Q >0$ (positive) refers to the heat absorbed
by the system and $\delta W >0$ (positive) refers to work performed by the system. 
Notice that, despite the energy is an observable and thus its differential $dE$ is exact,
those of the heat $Q$ and of the work $W$ are not.
For this reason, we indicate their infinitesimal variation with $\delta Q$ and $\delta W$.
While for classical systems, given a specific thermodynamic transformation, the heat and the work
are well defined quantities, for quantum systems this is not always the case and additional care
should be taken.

To be more quantitative, we start from the Hamiltonian $H = H(\lambda)$ of a generic quantum system,
which is supposed to depend on some macroscopic parameter $\lambda$
that can be suitably controlled (e.g., the magnetic field strength or the volume itself).
Its expectation value over the actual (generally mixed) state $\rho(\lambda)$
of the system defines the internal energy
\begin{equation}
  E \equiv \braket{H}_{\rho(\lambda)}= \text{Tr}\big[ H(\lambda) \,\rho(\lambda) \big],
\end{equation}
where ${\rm Tr}[\, \cdot \,]$ denotes the trace operation, so that
\begin{equation}
  dE  = \text{Tr} \big[ dH(\lambda) \, \rho(\lambda) \big] + \text{Tr} \big[ H(\lambda) \, d\rho(\lambda) \big].
  \label{eq:Emedio_generico}
\end{equation}
The first term in the r.h.s. accounts for a modification of the spectral structure of the system
as a response to a variation of $\lambda$,
while the second term describes a variation of the system state.
One could naively identify the first term as the work done by the system itself,
quantifying the mechanical energy exchanged with the external reservoir,
and the second one as the heat absorbed by the system from the environment~\cite{Alicki_1979, Kosloff1984, Quan_2007}.
However, this interpretation can be misleading, as, for example, when considering nonthermal reservoirs.  
In what follows, we consider the {\it heat} as the variation of energy due to the interaction
with a thermal reservoir and the {\it work} as the variation of energy due to a change in the Hamiltonian parameters. 

Once the heat exchange has been introduced, one can write
the Clausius inequality, 
\begin{equation}
  \sum_i \beta_i Q_i \le 0,
  \label{eq:clausius}
\end{equation}
where $Q_i$ denote the various heats exchanged with a set of surrounding reservoirs
at temperatures $\beta_i^{-1}$ (hereafter we will always work in units of $\hbar = k_B = 1$). 
Equation~\eqref{eq:clausius} can be combined with Eq.~\eqref{eq:1st_law} to show that,
for a single working medium operating between two reservoirs, cold (c) and hot (h),
at different temperatures and varying its energy,
only four operation modes are allowed~\cite{Solfanelli_2020}:
(i)   Heat Engine  (production of work by heat absorption from the hot reservoir, $Q_c<0, \, Q_h>0, \, W>0$);
(ii)  Refrigerator (heat transfer from the cold to the hot bath by energy absorption of the medium, $Q_c>0, \, Q_h<0, \, W<0$);
(iii) Thermal Accelerator (heat transfer from the hot to the cold bath by energy absorption, $Q_c<0, \, Q_h>0, \, W<0$);
(iv)  Heater (heating up the two baths by energy absorption, $Q_c<0, \, Q_h<0, \, W<0$).

\section{Thermodynamics of free-fermion systems}
\label{sec:td_fermionic_systems}

We now derive expressions for the work performed and the heat exchanged by free-fermion systems
during some paradigmatic and ideal thermodynamic transformations. 
Namely, we focus on perfect isotherm, adiabatic (unitary), and isochoric transformations. 

\subsection{Free-fermion systems}

We consider a lattice model of spinless fermions coupled through a generic quadratic Hamiltonian 
\begin{equation}
  H = \sum_{i,j} D_{i,j}  c_i^\dagger  c_j + \tfrac{1}{2} \big( O_{i,j} c_i^\dagger  c_j^\dagger + \text{h.c.} \big) ,
  \label{eq:fermionic_hamiltonian}
\end{equation}
where $c_i^{(\dagger)}$ are anticommuting fermionic annihilation (creation) operators
on the $i$-th site ($i=1,\ldots,N$), while $D$ and $O$ are complex $N \times N$ matrices
satisfying $D = D^\dagger$ and $O = - O^T$ (so to respect hermiticity of $H$).
By introducing the Nambu spinor $\Psi = \big( c_1, \dots,  c_N,  c_1^\dagger, \dots,  c_N^\dagger \big)^T$,
Eq.~\eqref{eq:fermionic_hamiltonian} can be written as
\begin{equation}
  H = \Psi^\dagger \mathbb{H} \Psi + \text{Tr}[D] ,
  \label{eq:H_nambu}
\end{equation}
where the $2N \times 2N$ matrix 
\begin{equation}
  \mathbb{H} = \frac{1}{2}\begin{pmatrix} \phantom{-}D & \phantom{-}O \\ -O^* & - D^* \end{pmatrix}
  \label{eq:bdg}
\end{equation}
is the so called Bogoliubov de-Gennes matrix, which can be diagonalized by a unitary transformation $\mathbb{U}$,
such that
\begin{equation}
  \mathbb{H^D} = \mathbb{U}^\dagger \mathbb{H} \mathbb{U}
  = \text{diag} \big\{ \epsilon_1 \ldots \epsilon_N, - \epsilon_1 \ldots -\epsilon_N \big\}.
\end{equation}
Here $\epsilon_k>0$ denote the positive eigenvalues of $\mathbb{H}$, $(k = 1, \dots, N)$,
while $\mathbb{U}$ is a Bogoliubov rotation that can be cast in the block form
\begin{equation}
  \mathbb{U} = \begin{pmatrix} \mathbb{u} & \mathbb{v}\\ \phantom{*}\mathbb{v}^* &\phantom{*} \mathbb{u}^* \end{pmatrix}.
\end{equation}
This transformation is constructed in such a way to preserve the particle-antiparticle symmetry of the model
and defines a new set of anticommuting fermions $b_k^{(\dagger)}$ (the Bogoliubov quasiparticles)
through the relation
$\Psi = \mathbb{U} \Phi$, with $\Phi = \big( b_1, \dots, b_N, b^\dagger_1, \dots, b_N^\dagger \big)^T$.
It is always possible to gauge away the term $\text{Tr}[D]$ in Eq.~\eqref{eq:H_nambu} 
by shifting the Hamiltonian, thus hereafter we simply neglect it.

Summarizing, in the diagonal basis, the Hamiltonian~\eqref{eq:fermionic_hamiltonian} reads
\begin{equation}
  H = \sum_{k>0} \omega_k(\lambda) \left( b^\dagger_k  b_k - \tfrac{1}{2}\right)
  \label{eq:H_bogo}
\end{equation}
and corresponds to a free-fermion model, being $\omega_k = 2 \epsilon_k$ the dispersion relation
of the (noninteracting) Bogoliubov $b_k$-quasiparticles~\cite{footnote:ksum}.
The spectrum of this Hamiltonian, which for our purposes we assume to be non degenerate,
should depend on the control parameter $\lambda$,
therefore we have $\omega_k \equiv \omega_k(\lambda)$.

\subsection{Ideal isotherm, adiabatic and isochoric transformations}
\label{sec:everything_ok}

For a quadratic fermionic system~\eqref{eq:fermionic_hamiltonian} prepared in the thermal state
\begin{equation}
  \rho = Z e^{-\beta  H},
  \label{eq:rho_therm}
\end{equation}
we can write the internal energy as
\begin{equation}
  E = \sum_k \omega_k(\lambda) \left(Z^{-1} \text{Tr}\big[ b^\dagger_k b_k e^{-\beta H} \big] - \tfrac{1}{2} \right),
  \label{eq:E1}
\end{equation}
where $\beta^{-1}$ denotes the temperature and $Z =\text{Tr}\big[e^{-\beta  H}\big]$ the partition function.
For a thermal density matrix as in Eq.~\eqref{eq:rho_therm},
the populations of the $b_k$ quasiparticles follow the Fermi-Dirac distribution function:
\begin{equation}
  \text{Tr} \big[ b^\dagger_k  b_k \, \rho \big] = \big[ 1 + e^{-\beta \omega_k(\lambda)} \big]^{-1}
  \equiv f\big[\beta, \omega_k(\lambda) \big],
\end{equation}
so that Eq.~\eqref{eq:E1} takes the simple form:
$E = \sum_k \omega_k(\lambda) \, \big\{ f\big[\beta, \omega_k(\lambda) \big] - 1/2 \big\}$.
The partition function can be easily computed to give:
\begin{equation}
    Z = e^{\frac{\beta}{2} \sum_k \omega_k(\lambda)} \prod_k \big[ 1 + e^{-\beta \omega_k(\lambda)} \big].
\end{equation}
The heat exchanged and the work performed by the system can be thus evaluated straightforwardly
in some paradigmatic transformations according to the following.

\subsubsection{Ideal isothermal transformation}

An ideal isotherm can be obtained by considering a slow variation of the control parameter $\lambda$,
such that the system is assumed to be always in thermal equilibrium with an environment
at a given fixed temperature $\beta^{-1}$. 
In that case, the Helmholtz free-energy is given by
\begin{equation}
  \mathcal{F} = - \frac{1}{\beta} \log Z = -\frac{\kappa}{2} - \frac{1}{\beta} \sum_k \log \big[ 1 + e^{-\beta \omega_k(\lambda)} \big],
\end{equation}
where $\kappa =  \sum_k \omega_k(\lambda)$.

The work performed by the system along an isothermal variation of the external parameter $\lambda$,
from $\lambda_i$ to $\lambda_f$, can be evaluated as the opposite of the variation
of the free-energy, that can be expressed in the integral form
\begin{align}
  \label{eq:work_qs}
  W = -\Delta \mathcal{F} & =
  -\int_{\lambda_i}^{\lambda_f} \frac{\partial \mathcal{F}}{\partial \lambda} d\lambda \\
  & = - \int_{\lambda_i}^{\lambda_f} \bigg\{ \sum_k \omega'_k(\lambda)
  \Big[ f\big[\beta,\omega_k(\lambda)\big] - \tfrac12 \Big] \bigg\} d\lambda , \nonumber \;
\end{align}
where the prime denotes the derivative with respect to $\lambda$.
Note that, depending on the derivative of spectrum $\omega_k'(\lambda)$,
the integrand in Eq.~\eqref{eq:work_qs} may behave monotonically with the temperature $\beta^{-1}$:
If $\omega_k'(\lambda) > 0$ $\forall k$, then the free-energy is an increasing function of $\beta$;
if $\omega_k'(\lambda) < 0$ $\forall k$, then the trend is reversed (details in App.~\ref{app:parametri_ising}).

The exchanged heat can be obtained through the first law of thermodynamics~\eqref{eq:1st_law},
$Q = \Delta E + W$, and using Eq.~\eqref{eq:E1}.

\subsubsection{Ideal adiabatic transformation}

By definition, an adiabatic transformation is such that it occurs without heat exchange ($Q = 0$).
We thus refer to this wording at any time the quantum system is isolated (not coupled to an external environment).
This has not to be confused with the quantum adiabaticity condition, i.e., when the unitary evolution
satisfies the requirements of the quantum adiabatic theorem~\cite{Sakurai_book} (thus a transformation can be adiabatic
but nonideal, in the sense that it is not quantum adiabatic --- see Sec.~\ref{subsec:real-adiabatic}).

Since, in general, an adiabatic transformation is not quasistatic, one cannot evaluate the work through the Helmholtz free-energy.
However, in the ideal case, the transformation is induced by a variation of $\lambda$,
from $\lambda_i$ to $\lambda_f$, which is so slow that one can invoke the quantum adiabatic
theorem~\cite{footnote:Adiab} (the spectrum is nondegenerate)
and calculate the work done by the system by exploiting the first law of thermodynamics:
\begin{equation}
  W = -\Delta E  = \big\langle H(\lambda_i) \big\rangle_{\rho(\lambda_i)} - \big\langle H(\lambda_f) \big\rangle_{\tilde{\rho}(\lambda_i)}.
\end{equation}
Here $\rho(\lambda_i)$ denotes the equilibrium thermal state for a system described by the Hamiltonian $H(\lambda_i)$
at temperature $\beta^{-1}$, at the beginning of the adiabatic transformation,
while $\tilde{\rho}(\lambda_i)$ is the quantum adiabatically evolved state.
Such latter state is constructed using the projectors on the eigenstates of $H(\lambda_f)$,
with the same energy-level populations of $\rho(\lambda_i)$, therefore: 
\begin{align}
  W &=\sum_k \big[ \omega_k(\lambda_i) - \omega_k(\lambda_f) \big] \Big\{ f\big[\beta,\omega_k(\lambda_i)\big] - \tfrac{1}{2}\Big\}, \nonumber \\
   & =- \int_{\lambda_i}^{\lambda_f} \bigg\{ \sum_k \omega'_k(\lambda)
  \Big[ f\big[\beta,\omega_k(\lambda_i)\big] - \tfrac12 \Big] \bigg\} d\lambda.
  \label{eq:work_ideal}
\end{align}

Note that in general, as in classical thermodynamics, the work done by the system
in an adiabatic transformation [cf.~Eq.~\eqref{eq:work_ideal}]
is not larger than the one done in an isothermal transformation [cf.~Eq.~\eqref{eq:work_qs}]. 
This follows from the fact that
\begin{equation}
  \int_{\lambda_i}^{\lambda_f}{\frac{ \omega_k'(\lambda)}{1 + e^{\beta \omega_k(\lambda)}}} \, d\lambda
  \leq \int_{\lambda_i}^{\lambda_f}{\frac{ \omega_k'(\lambda)}{1 + e^{\beta \omega_k(\lambda_i)}}} \, d\lambda,
\end{equation}
which can be proved by solving the integral and using the fact that $-\log\left(1 + e^{-x}\right)$ is a concave function.

\subsubsection{Ideal isochoric transformation}

An isochoric transformation is such that the system is put in contact with a thermal reservoir,
without variation in the Hamiltonian parameter $\lambda$.
No work is thus performed during this transformation ($W = 0$).
We model this situation in such a way that the system is prepared in a thermal state $\rho_i$
at temperature $\beta^{-1}_1$ and is put in contact with a bath at temperature $\beta_2^{-1}$,
keeping the Hamiltonian $H$ fixed.
After a time $\tau > t^\text{th}$, being $t^\text{th}$ the thermalization time, the system will be
in the thermal equilibrium state $\rho_f = e^{-\beta_2 H} / \text{Tr} [ e^{-\beta_2 H} ]$.
The heat exchanged during this process can be thus evaluated as the difference of Hamiltonian expectation values and reads 
\begin{equation}
  Q = \braket{H}_{\rho_f} - \braket{H}_{\rho_i} \\ 
  = \sum_k \omega_k\, \big[ f(\beta_2, \omega_k) - f(\beta_1, \omega_k) \big] . 
\end{equation}

In the next section, we discuss how to combine the latter two transformations to implement an Otto cycle and the generalization of these results when the assumptions of perfect thermalization and quantum adiabaticity are relaxed.

\section{The quantum Otto cycle}
\label{sec:The quantum Otto cycle}

In this section we derive the expression for the work performed by a real quantum Otto engine, i.e.,
an engine whose transformations may violate the assumptions of perfect adiabaticity and/or perfect thermalization. 
We consider, as a working substance (medium), a system of coupled fermions,
according to Eq.~\eqref{eq:fermionic_hamiltonian}. After a Bogoliubov transformation,
such Hamiltonian can be mapped into the free-quasiparticle model~\eqref{eq:H_bogo},
with dispersion relation $\omega_k(\lambda)$. As we shall see in a moment, the only ingredient needed
to assess the performance of such an engine is the quasiparticles distribution on the evolved system state.

The quantum Otto cycle operates between the two temperatures $\beta_c^{-1}$ (cold) and $\beta_h^{-1}$ (hot),
and consists of four strokes: two (thermodynamic) adiabatic transformations (i.e., no heat exchange with the environment)
and two isochoric transformations (i.e., no work on the system)~\cite{Henrich2007}.
The system, initially prepared in the thermal state $\rho_c$ at temperature $\beta_c^{-1}$, undergoes the following steps:
\begin{enumerate}
\item $[$Forward adiabatic transformation$]$: The Hamiltonian parameter $\lambda$ is varied
  in a time $T$, from $\lambda_i$ to $\lambda_f$; \label{aa}
\item $[$Hot isochoric thermalization$]$: The Hamiltonian $H_f \equiv  H(\lambda_f)$ is fixed and the system
  is put in contact with the reservoir at temperature $\beta_h^{-1} > \beta_c^{-1}$, for a time $\tau$;
\item $[$Backward adiabatic transformation$]$: The Hamiltonian parameter $\lambda$ is varied,
  in a time $T$ from $\lambda_f$ to $\lambda_i$, thus reversing point~\ref{aa};
\item $[$Cold isochoric thermalization$]$: The Hamiltonian $H_i \equiv H(\lambda_i)$ is fixed and the system
  is put in contact with the reservoir at temperature $\beta_c^{-1} < \beta_h^{-1}$ for a time $\tau$.
\end{enumerate}

\subsection{Ideal free-fermion cycle}

We consider, as a working medium, a quadratic fermionic system weakly coupled to two thermal reservoirs,
as modeled in App.~\ref{app:thermalization}.
By exploiting the results of Sec.~\ref{sec:everything_ok}, the work performed and heat exchanged
for an ideal Otto cycle (strictly speaking, perfect thermalization and perfect quantum adiabatic sweeps are ideally obtained
in the limits $T\to \infty$, $\tau \to \infty$) can be written as: 
\begin{subequations}
  \label{eq:heat}
  \begin{align}
    Q_h^\text{id} = & \phantom{-} \sum_k \omega_k(\lambda_f) \, \big( \Delta f_k \big)_{hc} \,, \\
    Q_c^\text{id} = &          -  \sum_k \omega_k(\lambda_i) \, \big( \Delta f_k \big)_{hc}\,, \\
    W^\text{id}   = & \phantom{-} \sum_k \big[\omega_k(\lambda_f) - \omega_k(\lambda_i) \big] \, \big( \Delta f_k \big)_{hc} \,,\quad
  \end{align}
  where
\begin{equation}
  \big( \Delta f_k \big)_{hc} \equiv f \big[ \beta_h, \omega_k(\lambda_f) \big] - f \big[ \beta_c, \omega_k(\lambda_i) \big].
  \end{equation}
\end{subequations}
Note that, since after a single ideal cycle the system comes back to the same initial state,
we have $\Delta E=0$ and thus $W^\text{id} = Q^\text{id}_h +Q^\text{id}_c$ follows from the first principle of thermodynamics.
The efficiency of the cycle can be thus expressed as:
\begin{equation}
  \eta^\text{id} \equiv \frac{W^\text{id}}{Q_h^\text{id}}
  = 1 - \frac{\sum_k \omega_k(\lambda_i)\big( \Delta f_k \big)_{hc}}{\sum_q \omega_q(\lambda_f)\big( \Delta f_q \big)_{hc} }.
  \label{eq:efficeincy}
\end{equation}

We can obtain some basic considerations by introducing the ratio
\begin{equation}
  r_k(\lambda_i,\lambda_f) \equiv \omega_k(\lambda_i) / \omega_k(\lambda_f).
  \label{eq:ratio}
\end{equation}
In fact, the sign of $\left(\Delta f_k\right)_{hc}$ depends on the value assumed by $r_k(\lambda_i,\lambda_f)$.
If $r_k(\lambda_i,\lambda_f) > \beta_h/\beta_c$, then $\left(\Delta f_k\right)_{hc}$ is positive;
otherwise it is negative. Since we have $\omega_k(\lambda) > 0 \hspace{0.2cm} \forall k$, from the sign of $\left(\Delta f_k\right)_{hc}$ we can infer some information on the way the engine is operating. In particular,
\begin{itemize}
\item  If $r_k(\lambda_i,\lambda_f) > \beta_h/\beta_c \  \forall k$, then $Q_h > 0$ and $Q_c < 0$,
  thus the engine works either as a thermal accelerator (if $W<0$) or as a heat engine (if $W>0$);
\item  If $r_k(\lambda_i,\lambda_f) > 1 \  \forall k$, then $Q_h > 0$, $Q_c < 0$, and $W < 0$,
  thus the engine is a thermal accelerator;
\item  If $\beta_h / \beta_c < r_k(\lambda_i,\lambda_f) < 1 \  \forall k$, then $Q_h > 0$, $Q_c < 0$, and $W > 0$,
  thus the engine is a heat engine;
\item If $r_k(\lambda_i,\lambda_f) < \beta_h/\beta_c \ \forall k$, then $Q_h < 0$ and $Q_c > 0$,
  thus the engine is a refrigerator. 
 \label{item:dispersionrelation}
\end{itemize}
These conditions are valid for any system described by a quadratic fermionic Hamiltonian.
It is important to remark that they are sufficient, but not necessary, conditions
[e.g., if $r_k(\lambda_i,\lambda_f) > \beta_h/\beta_c$ only for some $k$, it is still possible to 
have an engine operating as a thermal accelerator or as a heat engine].

\subsection{Nonperfect thermalization}
\label{subsec:non-therm}

We now relax the hypothesis of perfect thermalization, by assuming the system to be in contact
with the two reservoirs for a time $\tau < t^\text{th}$, where $t^\text{th}$ is the thermalization time. 
According to Ref.~\cite{DAbbruzzo_2021} the nonperfect thermalized distribution of the quasiparticles reads
\begin{equation}
  \braket{ b^\dagger_k  b_k}(\gamma) = f(\beta, \omega_k)(1 - e^{- \mathcal{J} \gamma})
  + \braket{ b^\dagger_k  b_k}_{\rho_0} e^{- \mathcal{J} \gamma},
  \label{eq:correlator_nonth}
\end{equation}
where we have put $\gamma \equiv 2 \tau$ and hereafter we assume $\mathcal{J}=1$.
Note that $\mathcal{J}$ is a parameter which depends on the bath properties and on the system-bath coupling,
as detailed in App.~\ref{app:thermalization}, and it obviously affects the time scale of thermalization.
The off diagonal correlators and the anomalous ones $\braket{ b_k  b_q}$ remain zero at any time,
provided the initial condition for the Otto cycle is a thermal state (say, $\rho_c$).
Equation~\eqref{eq:correlator_nonth} ensures that convergence to the perfect thermal behavior
is exponential although, strictly speaking, perfect thermalization cannot be reached
at any finite time ($t^\text{th} \to \infty$).
Thus the ideal case would correspond to the limit $\tau \to \infty$.

Let us find an expression for the asymptotic distribution of quasiparticles after $n$ repetitions
of the nonperfect thermalized cycle are implemented. 
We introduce the diagonal $N \times N$ matrices~\cite{footnote:Diag}
\begin{eqnarray}
  \mathbb{\Theta}_{c(h)} & = & \text{diag} \big\{ f [\beta_{c(h)}, \omega_k(\lambda_{i(f)})] \big\}_{k=1,\ldots,N} \: , \\
  \mathbb{\Gamma}_{c(h)}^{[n]} & = & \text{diag} \big\{ \text{Tr} [ b^\dagger_k  b_k \, \tilde{\rho}_{c(h)}^{[n]} ] \big\}_{k=1,\ldots,N} \: , \label{eq:gamma}
\end{eqnarray}
where $\tilde{\rho}_{c(h)}^{[n]}$ denotes the state at the end of the nonperfect thermalized stroke
with the bath at temperature $\beta_{c(h)}^{-1}$ of the $n$-th cycle repetition.
The system is initially prepared in the thermal state $\rho_c$ at temperature $\beta^{-1}_c$, i.e. $\mathbb{\Gamma}^{[0]}_c \equiv \mathbb{\Theta}_c$. Since the asymptotic distribution does not depend on the initial condition, this choice does not affect the generality of the result.
Then, at the end of the ideal adiabatic transformation, the system is put in contact with the reservoir at temperature $\beta_h^{-1}$ and reaches the state $\rho_h^{[1]}$, characterized by the distribution
\begin{subequations}
  \label{eq:Nonth1}
  \begin{equation}
    \mathbb{\Gamma}_h^{[1]} = \mathbb{\Theta}_h(1 - e^{-\gamma}) + \mathbb{\Gamma}^{[0]}_c e^{-\gamma}.
  \end{equation}
  After the backward ideal adiabatic transformation is implemented, the system is put in contact with the reservoir at temperature $\beta_c^{-1}$ reaching the state $\rho_c^{[1]}$, characterized by the distribution
  \begin{equation}
    \mathbb{\Gamma}^{[1]}_c = \mathbb{\Theta}_c(1 - e^{-\gamma}) + \mathbb{\Gamma}_h^{[1]} e^{-\gamma}.
  \end{equation}
\end{subequations}
The two equations~\eqref{eq:Nonth1} can be easily generalized to the $n$-th iteration,
to obtain
\begin{subequations}
  \begin{align}
    &\mathbb{\Gamma}_c^{[n]} = \big(\mathbb{\Theta}_c + e^{-\gamma} \mathbb{\Theta}_h\big) (1 - e^{-\gamma}) + \mathbb{\Gamma}_c^{[n-1]} e^{-2\gamma},\\ 
    &\mathbb{\Gamma}_h^{[n]} = \big(\mathbb{\Theta}_h + e^{-\gamma} \mathbb{\Theta}_c\big) (1 - e^{-\gamma}) + \mathbb{\Gamma}_h^{[n-1]} e^{-2\gamma}.
  \end{align}
\end{subequations}
By imposing the stationary condition that both the $\mathbb{\Gamma}$'s on the left and right hand side are iteration independent,
one gets the stationary solution
\begin{subequations}
  \label{eq:fixed_adiabatic}
  \begin{align}
    &\mathbb{\Gamma}_c^{\infty} = h(\gamma)\big(\mathbb{\Theta}_c + e^{-\gamma} \mathbb{\Theta}_h\big), \\ 
    &\mathbb{\Gamma}_h^{\infty} = h(\gamma) \big(\mathbb{\Theta}_h + e^{-\gamma} \mathbb{\Theta}_c\big),
  \end{align}
\end{subequations}
where $h(\gamma) = (1 + e^{-\gamma})^{-1}$. 
Being 
\begin{equation}
	\frac{d \big(\mathbb{\Gamma}_i^{[n]}\big)_{kk}}{d\big(\mathbb{\Gamma}_i^{[n-1]}\big)_{kk}} = e^{-2\gamma} < 1,
\end{equation}
with $\big(\mathbb{\Gamma}_i^{[n]}\big)_{kk}$ the $k$-th diagonal element of the matrix ($i=c,h$),
the fixed points in Eq.~\eqref{eq:fixed_adiabatic} are stable and the convergence is exponential in $n$.

Equations~\eqref{eq:fixed_adiabatic} suggest that, for the nonperfectly thermalized cycle (n-th),
the heat absorbed and the work performed per cycle are evaluated from Eq.~\eqref{eq:heat}
by substituting the rescaled Fermi functions 
\begin{eqnarray}
    f[\beta_{c}, \omega_k(\lambda_i)] & \!\mapsto \!& h(\gamma)\big\{ f[\beta_{c}, \omega_k(\lambda_i)] + e^{-\gamma} f[\beta_{h}, \omega_k(\lambda_f)] \big\}, \nonumber \\
    f[\beta_{h}, \omega_k(\lambda_f)] & \!\mapsto \!& h(\gamma)\big\{ f[\beta_{h}, \omega_k(\lambda_f)] + e^{-\gamma} f[\beta_{c}, \omega_k(\lambda_i)] \big\},  \nonumber
\end{eqnarray}
thus obtaining
\begin{equation}
  Q_{h(c)}^\text{n-th} = g(\gamma) \, Q_{h(c)}^\text{id}, \qquad W^\text{n-th} = g(\gamma) \, W^\text{id},  
\end{equation}
where $g(\gamma ) = (1 - e^{-\gamma})h(\gamma) = \tanh(\gamma/2)$.
Therefore it follows that, in this case, the efficiency of the heat engine
$\eta^\text{n-th} \equiv W^\text{n-th} / Q_h^\text{id}$
remains the same as for the ideal case [see Eq.~\eqref{eq:efficeincy}].

\begin{figure}
  \centering
  \includegraphics[width=0.49\textwidth]{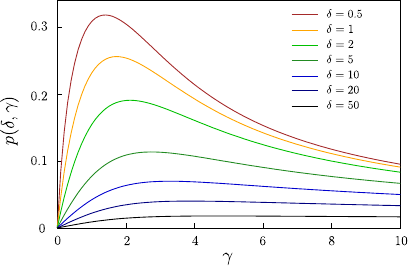}
  \caption{The value of $p(\delta, \gamma)$ versus the thermalization time $\gamma= 2\tau$,
    for different values of the quench duration $\delta$ (color scale in the legend).}
  \label{fig:p_delta_eta}
\end{figure}

\begin{figure}
  \centering
  \includegraphics[width=0.49\textwidth]{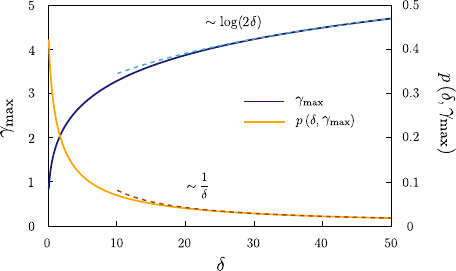}
  \caption{The value of $\gamma_\text{max}$ that maximizes the power output (blue curve) and the corresponding
    value of $p(\delta, \gamma_\text{max})$ (orange curve) versus $\delta$.
    Dashed lines denote the asymptotic large-$\delta$ behaviors analytically evaluated.}
  \label{fig:max_nth}
\end{figure}

If $W > 0$ we  can define the power of the engine as function of the thermalization time ($\tau = \gamma/2$)
and of the quench time ($T = \delta/2$):
\begin{equation}
  \mathcal{P}^\text{n-th}(\delta, \gamma) \equiv \frac{W^\text{n-th}}{\delta + \gamma} = p(\delta, \gamma) \, W^\text{id}, 
  \label{eq:potenzanonterm}
\end{equation}
with
\begin{equation} 
     p(\delta, \gamma) \equiv \frac{\tanh(\gamma/2)}{\delta + \gamma}. 
\end{equation}
In this case we are not concerning about nonadiabatic effects, i.e., Eq.~\eqref{eq:potenzanonterm} is valid if we choose $\delta$ such that, 
according to the features of the quenches ($\mathbb{H}(\lambda)$, $\lambda_i$, $\lambda_f$),
the quantum adiabatic theorem approximately holds (see Sec.~\ref{subsec:real-adiabatic}).   
Of course, this value of $\delta$ is model dependent.
In Fig.~\ref{fig:p_delta_eta} we show the behavior of $p(\delta, \gamma)$ versus $\gamma$, for different values of $\delta$ fixed.
Since the work for the ideal case $W^\text{id}$ is independent of $\delta$ and $\gamma$,
the power is maximum for cycle parameters maximizing $p(\delta, \gamma).$
Therefore, we can optimize the performance by fixing $\delta$ (and assuming that it is suitable for adiabatic theorem to hold) and maximizing with respect to $\gamma$,
in order to have some indication on the best $\gamma$ for any given quench duration. The maximization
\begin{equation}
  \partial_\gamma p(\delta, \gamma) = \frac{1}{\delta + \gamma} \bigg[ \frac{1}{2 \cosh^2 (\gamma/2)}- p(\delta, \gamma) \bigg] = 0
\end{equation}
leads to the stationarity condition
\begin{equation}
  \frac{e^{\gamma} - e^{-\gamma}}{2} - \gamma = \delta.
\end{equation}
Let us define $S(\gamma) = \frac{e^{\gamma} - e^{-\gamma}}{2} - \gamma$.
For $\gamma \gg 1$, $S(\gamma) \sim e^{\gamma}/2$. As a consequence, $\gamma_\text{max} \sim \log(2\delta)$, i.e.,
the maximum scales logarithmically with the quench duration. 
In correspondence of the maximum we have
\begin{equation}
  p \big( \delta, \log(2\delta) \big) \sim
  \frac{1}{\delta} \frac{(2\delta)^{1/2} - (2\delta)^{-1/2}}{(2\delta)^{1/2} + (2\delta)^{-1/2}}.
\end{equation}
Therefore, for $\delta \to \infty$, we have $p(\delta, \gamma) \sim 1/\delta \to 0$, i.e., no power is produced by the engine.  
We observe that for $\gamma \ll 1$ we have $S(\gamma) \sim \gamma^3/6$, leading to $\gamma_\text{max} \sim (6\delta)^{1/3}$,
with $\delta \ll 1$. By substituting this value, we notice that for fast adiabatic sweeps
we have $p(\delta, \gamma_\text{max}) \to 0.5$, paying attention that this result holds for regimes
in which even for small $\delta$ quantum adiabaticity is still valid.
In Fig.~\ref{fig:max_nth} we show $\gamma_\text{max}$ (blue curve) and $p(\delta, \gamma_\text{max})$ (orange curve) vs $\delta$.
The dashed lines are the expected analytical results for large $\delta$.

\subsection{Real adiabatic processes}
\label{subsec:real-adiabatic}

We now consider a system prepared in the thermal state $\rho_1$ with an Hamiltonian $H_i$
characterized by a spectrum $\omega_k(\lambda_i)$. We implement a transformation $\lambda_i \to \lambda_f$
in a time $T$ such that the quantum adiabatic regime is no more valid~\cite{footnote:Adiab}.
As discussed in Sec.~\ref{sec:everything_ok}, the work can be evaluated from the first law of thermodynamics. 
However, because of the nonperfect quantum adiabaticity of the transformation, some excitations are generated during the dynamics.

The unitary dynamics is generated by the operator $U_{ev}(t) = \mathcal{T}\exp \big[ e^{-i\int_0^t dsH(s)} \big]$, so that
\begin{equation}
  W = \ \braket{ H_i}_{\rho_1} - \braket{ H_f}_{\rho_1(T)},
  \label{eq:expressionrealwork}
\end{equation}
where $\rho_1(T) =  U_\text{ev}(T) \, \rho_1 \, U_\text{ev}^\dagger(T)$ denotes the post-quench state.
The first term in the r.h.s.~is known, while the second one can be easily evaluated
in the Nambu spinor notation, using Eq.~\eqref{eq:H_nambu}:
\begin{align}
  \braket{H_f}_{\rho_1(T)} = 
  & \sum_{i,j}\text{Tr} \big[ \Psi^\dagger_i (\mathbb{H}_{f})_{ij} \Psi_j \, U_\text{ev}(T) \rho_1 U_\text{ev}^\dagger(T) \big] \nonumber\\
  = & \sum_{i,j}\text{Tr} \big[ \Psi^{H \dagger}_i (\mathbb{H}_{f})_{ij}\Psi^H_j \rho_1\big],
  \label{eq:work_real}
\end{align}
where we adopted the Heisenberg representation and introduced the time-evolved operators 
\begin{align}
  \Psi^H & = U_\text{ev}^\dagger(T) \Psi U_\text{ev} (T) \nonumber \\
  & = \big( c_1^{H}(T), \dots,  c_N^{H}(T),  c_1^{H\dagger}(T), \dots,  c_N^{H\dagger}(T) \big)^T.
\label{eq:b_time_evolved}
\end{align}

As discussed in Refs.~\cite{Mbeng_2020, Piccitto_2023}, these operators can be expressed
in terms of the $ b_k$-fermions by implementing the following time-dependent Bogoliubov transformation
\begin{equation}
	\Psi^H = \mathbb{V}(T) \Phi,
\end{equation}
where $\mathbb{V}(T)$ is a matrix obeying the following equation
\begin{equation}
  \partial_s \mathbb{V}(s) = -2i\mathbb{H}(s) \mathbb{V}(s),
  \label{eq:Vevol}
\end{equation}
with initial condition $\mathbb{V}(0) = \mathbb{U}(h_i)$. 
Substituting this result in Eq.~\eqref{eq:work_real}, we get
\begin{equation}
	\braket{ H_f}_{\rho_1(T)} = \sum_{i,j}\text{Tr}\big[\Phi_i^\dagger \mathbb{\tilde{H}}_{ij} \Phi_j \rho_1\big],
\end{equation}
being $\mathbb{\tilde{H}} = \mathbb{V}^\dagger(T) \mathbb{H}_{f}  \mathbb{V}(T)$.
By construction, the state $\rho_1$ is diagonal in the $b_k$-fermions (i.e. $\braket{ b_q^\dagger  b_{q'}} = 0$, if $q \neq q'$), therefore we have
\begin{equation}
  \braket{H_f}_{\rho_1(T)} = \sum_k \tilde{\omega}_k \Big\{ f\big[ \beta, \omega_k(\lambda_i) \big] - \tfrac{1}{2} \Big\},
  \label{eq:work_real_finale}
\end{equation}
where $\tilde{\omega}_k \equiv \mathbb{\tilde{H}}_{kk}$ (for $1 \leq k \leq N$) and we used the fact that $\mathbb{V}(T)$
is a Bogoliubov transformation and so it maintains particle-hole symmetry
$(\mathbb{\tilde{H}}_{kk} = -\mathbb{\tilde{H}}_{k+N,k+N})$.
Collecting all these results, the work at the end of the adiabatic stroke reads
\begin{equation}
  W = \sum_k \big[ \omega_k(\lambda_i) - \tilde{\omega}_k(\lambda_f) \big]
  \Big\{ f[\beta,\omega_k(\lambda_i)]-\tfrac{1}{2} \Big\}.
\end{equation}

From a different perspective, in the Heisenberg picture, we can introduce a new set of fermions $ d_k^{(\dagger)} \equiv b_k^{H(\dagger)}$, whose Nambu spinor $\Xi = ( d_1, \dots,  d_N,  d_1^\dagger, \dots,  d_N^\dagger)^T$ relates to the quasiparticles one through the Bogoliubov transformation $\mathbb{Q} = \mathbb{U}^{\dagger}(\lambda_f) \mathbb{V}(T)$, $\Xi = \mathbb{Q} \Phi$. In this way, we can write 
\begin{equation}
\braket{ \Phi^{\dagger}_k  \Phi_q}_{\rho_1(T)} =\braket{ \Xi_k^{\dagger}  \Xi_q}_{\rho_1} = \sum_{i,j=1}^{2N}\mathbb{Q}^{*}_{ki}\braket{ \Phi_i^{\dagger}  \Phi_j}_{\rho_1}\mathbb{Q}_{qj} .
\end{equation}

Note that, in the limit of an ideal adiabatic transformation, the unitary evolution of the eigenvectors of
$\mathbb{H}(\lambda_0)$ [Eq.~\eqref{eq:Vevol}] is such that these ones are instantaneous
eigenvectors of $\mathbb{H}(t)$, therefore $\mathbb{V}^\dagger(T) \, \mathbb{H}_{f} \, \mathbb{V}(T) =
\mathbb{U}^\dagger(\lambda_f) \, \mathbb{H}_f \, \mathbb{U}(\lambda_f) = \mathbb{H^D}_f$.
So $\mathbb{\tilde{H}} = \mathbb{H^D}_f$ and Eq.~\eqref{eq:expressionrealwork} reduces to Eq.~\eqref{eq:work_ideal}.

We also point out that, as shown in Ref.~\cite{allahverdyan2004maximal}, imperfect quantum adiabatic transformations
starting from a thermal state excite the system more than the ideal quantum adiabatic ones.
As a consequence, the former are expected to produce less work than latter case.
In fact, we verified such statement through several numerical simulations, always obtaining
$\tilde{\omega}_{k} < \omega_{k}(\lambda_f)$, that is $W^{id} > W$.

\subsection{Nonperfect thermalization and real adiabatic processes}
\label{subsec:non-thermal+real_adiabatic}

Let us finally assume nonperfect thermalization and real adiabatic transformations.
The system is prepared in the thermal state at temperature $\beta_1^{-1}$, $\rho_0 \equiv \rho_1$. 
We can follow a procedure similar to that of Sec.~\ref{subsec:non-therm}, but we have to identify two other
intermediate steps in the dynamics. 
Moreover, in the case of real adiabatic transformations, the nondiagonal components of the correlator
of the Bogoliubov quasiparticles can be nonvanishing. 
To account for this effect, we derive the results using the Nambu spinors
and, in analogy with Eq.~\eqref{eq:gamma}, define
\begin{equation}
  \big(\mathbb{\Lambda}_i^{[n]}\big)_{jl} = \braket{\Phi_j \Phi^\dagger_l}_{\rho_i^{[n]}}, 
\end{equation}
where $\rho_i^{[n]}$ represents the state of the system after the $n$th thermalization with the bath at temperature $\beta_i^{-1}$.
We also define
\begin{equation}
	\big(\mathbb{\Lambda}_{i, T}^{[n]}\big)_{jl} = \braket{\Phi_j \Phi^\dagger_l}_{\rho_i^{[n]}(T)},
\end{equation}
being $\rho_i^{[n]}(T) = \mathbb{U}(T) \rho_i^{[n]} \mathbb{U}(T)^\dagger$ the time evolved density matrix. 
We notice that
\begin{equation}
  \begin{aligned}
    \braket{\Phi_j \Phi^\dagger_l}_{\rho_i^{[n]}(T)} = \ &\text{Tr}\big[ \Phi_j\Phi^{\dagger}_l \rho_1^{[n]}(T)\big] \\
    =\ &\text{Tr}\big[ \Xi_j\Xi_l^{\dagger} \,\rho_1^{[n]}\big] \\
    =\ &\text{Tr}\sum_{mn}\big[ \mathbb{Q}_{jm} \Phi_m\Phi^{\dagger}_n \,\mathbb{Q}^{\dagger}_{nl}\rho_1^{[n]}\big] \\
    =\ &\sum_{mn}\mathbb{Q}_{jm} \text{Tr}\big[ \Phi_m\Phi_n^{\dagger}\,\rho_1^{[n]}\big]  \mathbb{Q}^{\dagger}_{nl}\\ 
    =\ & \Big(\mathbb{Q} \mathbb{\Lambda}_i^{[n]} \mathbb{Q}^\dagger\Big)_{jl}.
  \end{aligned}
\end{equation}
We introduce the diagonal $2N \times 2N$ matrix
\begin{equation}
  \mathbb{\Omega}_{c(h)} = \text{diag} \big\{ f \left[ \beta_{c(h)}, -\omega_k(\lambda_{i(f)}) \right] ,
  f \left[ \beta_{c(h)}, \omega_k(\lambda_{i(f)}) \right] \big\},
\end{equation}
with $i = c, h$.
We start in the thermal state $\mathbb{\Lambda}^{[0]}_1 = \mathbb{\Omega}_c$.

After the first adiabatic stroke we have
\begin{equation}
	\mathbb{\Lambda}^{[1]}_{1,T} = \mathbb{Q} \mathbb{\Lambda}_1^{[0]} \mathbb{Q}^\dagger    
\end{equation}
Then the system is put in contact with the reservoir at temperature $\beta_2^{-1}$ for a time $\tau$.
Differently from Sec.~\ref{subsec:non-therm}, in this case the system at the end of the adiabatic sweep
could present some excitations, so we have to consider also the evolution of nondiagonal elements
of the correlation functions~\cite{DAbbruzzo_2021}
\begin{subequations}
  \label{eq:Bath_corr}
  \begin{eqnarray}
 \braket{b^{\dagger}_k b_q }(\tau) & = & \braket{ b^{\dagger}_k b_q }_{\rho_0} e^{i(\omega_k - \omega_q)\tau -\gamma},\\
    \braket{ b^{\dagger}_k b^{\dagger}_q }(\tau) & = & \braket{ b^{\dagger}_k b^{\dagger}_q }_{\rho_0} e^{i (\omega_k + \omega_q)\tau -\gamma },\\
   \braket{ b_k b_q }(\tau) & = & \braket{ b_k b_q }_{\rho_0} e^{-i(\omega_k + \omega_q)\tau -\gamma}.  
\end{eqnarray}
\end{subequations}
At the end of this process we have
\begin{equation}
	\mathbb{\Lambda}^{[1]}_2 = \mathbb{\Omega}_h (1 - e^{-\gamma}) + \mathbb{\Phi}\mathbb{\Lambda}^{[1]}_{1, T}\mathbb{\Phi}^{\dagger} e^{-\gamma},
\end{equation}
where
\begin{equation}
  \mathbb{\Phi} \!\equiv \!\text{diag} \big[ e^{-i\omega_1\!(h_i)\tau}, \ldots, e^{-i\omega_N\!(h_i)\tau},
    e^{i\omega_1\!(h_i)\tau}, \ldots, e^{i\omega_N\!(h_i)\tau} \big].
\end{equation}
We perform the adiabatic transformation ending with
\begin{equation}
  \mathbb{\Lambda}^{[1]}_{2,T} = \mathbb{Q'} \Lambda_1^{[0]} \mathbb{Q'}^\dagger,    
\end{equation}
being $\mathbb{Q'}$ the rotation diagonalizing the Hamiltonian after the backward adiabatic transformation.
Finally we connect the system to the reservoir at temperature $\beta_1^{-1}$ obtaining
\begin{equation}
	\mathbb{\Lambda}^{[1]}_1 = \mathbb{\Omega}_c (1 - e^{-\gamma}) + \mathbb{\Phi}'\mathbb{\Lambda}^{[1]}_{2,T}\mathbb{\Phi}'^{\dagger} e^{-\gamma},
\end{equation}
where $\mathbb{\Phi}'$ is the same as $\mathbb{\Phi}$, but with the final eigenstates $\omega_k(h_f)$.
In what follows, to simplify the notation, we redefine the matrices $\mathbb{Q}$, $\mathbb{Q}'$
including $\mathbb{\Phi}$, $\mathbb{\Phi}'$ in their definition
[i.e., $\mathbb{Q} = \mathbb{\Phi} \, \mathbb{U}^{\dagger}(h_f) \, \mathbb{V}(T)$].
This does not change the mean value of the Hamiltonian at the end of the adiabatic sweeps.
At the $n$-th iteration we find 
\begin{subequations}
  \begin{eqnarray}
    \mathbb{\Lambda}^{[n]}_{1,T} & = & \mathbb{Q}\mathbb{\Lambda}^{[n-1]}_1\mathbb{Q}^\dagger ,\\
    \mathbb{\Lambda}^{[n]}_2    & = & \mathbb{\Omega}_h (1 - e^{-\gamma}) + \mathbb{\Lambda}^{[n]}_{1,T} e^{-\gamma}, \\
    \mathbb{\Lambda}^{[n]}_{2,T} & = & \mathbb{Q'}\mathbb{\Lambda}^{[n]}_2\mathbb{Q'^{\dagger}},\\
    \mathbb{\Lambda}^{[n]}_1    & = & \mathbb{\Omega}_c (1 - e^{-\gamma}) + \mathbb{\Lambda}^{[n]}_{2,T} e^{-\gamma}.
  \end{eqnarray}
\end{subequations}
Substituting we obtain 
\begin{subequations}
  \label{eq:sistema_molto_brutto}
  \begin{eqnarray}
    \mathbb{\Lambda}^{[n]}_{1,T} & = & \mathbb{Q} \mathbb{K}_1 \mathbb{Q}^\dagger + e^{-2\gamma} \mathbb{Q}\mathbb{Q'}\mathbb{\Lambda}_{1,T}^{[n-1]}\mathbb{Q'}^\dagger \mathbb{Q}^\dagger, \\
    \mathbb{\Lambda}^{[n]}_2 & =& \mathbb{K}_2 + e^{-2\gamma} \mathbb{Q}\mathbb{Q'}\mathbb{\Lambda}_{2}^{[n-1]}\mathbb{Q'}^\dagger \mathbb{Q}^\dagger, \\
    \mathbb{\Lambda}^{[n]}_{2,T} & = & \mathbb{Q}'\mathbb{K}_2 \mathbb{Q}'^\dagger + e^{-2\gamma} \mathbb{Q'}\mathbb{Q}\mathbb{\Lambda}_{2,T}^{[n-1]}\mathbb{Q}^\dagger \mathbb{Q'}^\dagger, \\
    \mathbb{\Lambda}^{[n]}_1 & = & \mathbb{K}_1 + e^{-2\gamma} \mathbb{Q'}\mathbb{Q}\mathbb{\Lambda}_{1}^{[n-1]}\mathbb{Q}^\dagger \mathbb{Q'}^\dagger,
  \end{eqnarray}
\end{subequations}
having introduced 
\begin{subequations}
  \begin{eqnarray}
    \mathbb{K}_1 & = & (1 - e^{-\gamma})(\mathbb{\Omega}_c + e^{-\gamma} \mathbb{Q}'\mathbb{\Omega}_h\mathbb{Q}'^\dagger),\\
    \mathbb{K}_2 & = & (1 - e^{-\gamma})(\mathbb{\Omega}_h + e^{-\gamma} \mathbb{Q}\mathbb{\Omega}_c\mathbb{Q}^\dagger).
  \end{eqnarray}
\end{subequations}
We can write explicitly the series in the first equation of dynamical systems Eq.~\eqref{eq:sistema_molto_brutto}, finding
\begin{align}
    \mathbb{\Lambda}^{[n+1]}_{1,T} = & \sum_{k=0}^{n-1} e^{-2k\gamma} \big[ (\mathbb{Q}\mathbb{Q'})^k \mathbb{Q}\mathbb{K}_{1}\mathbb{Q}^\dagger (\mathbb{Q'}^\dagger \mathbb{Q}^\dagger)^k \big] \nonumber \\ 
    & + e^{-2n \gamma } \big[ (\mathbb{Q}\mathbb{Q'})^{n} \mathbb{\Lambda}^{[1]}_{i, T} (\mathbb{Q'}^\dagger \mathbb{Q}^\dagger)^{n} \big].
\end{align}
The asymptotic solution of this equation reads
\begin{equation}
  \mathbb{\Lambda}^{\infty}_{1, T} = \sum_{k=0}^{\infty} e^{-2k\gamma} \big[ (\mathbb{Q}\mathbb{Q'})^k \mathbb{Q}\mathbb{K}_{1} \mathbb{Q}^\dagger (\mathbb{Q'}^\dagger \mathbb{Q}^\dagger)^k \big].
  \label{eq:lambda_infinito}
\end{equation}
Similar expressions can be written for the other matrices. 

Before concluding we note that, due to the unitarity of the matrices $\mathbb{Q}, \mathbb{Q'}$,
and the fact that $\mathbb{\Omega}_{c(h)}$ is a diagonal matrix with real elements in $[0,1]$, we have:
\begin{equation}
  \Big\lvert \big[ (\mathbb{Q}\mathbb{Q'})^k \mathbb{Q}\mathbb{K}_{1} \mathbb{Q}^\dagger (\mathbb{Q'}^\dagger \mathbb{Q}^\dagger)^k\big]_{ij}\Big\rvert <\left(1 - e^{-2\gamma}\right).
\end{equation}
This means that the terms in Eq.~\eqref{eq:lambda_infinito} decay exponentially to zero with $k$.
In particular, the modulus of each term is smaller than 
\begin{equation}
  \xi(k) \equiv e^{-2k\gamma}\left(1 - e^{-2\gamma}\right).
\label{eq:error}
\end{equation}
So we can safely approximate the asymptotic value of $\mathbb{\Lambda}_1^\infty$ with the first $K$ terms of the series. 
The convergence of $\mathbb{\Lambda}^{n}_{1, T}$ to $\mathbb{\Lambda}^{\infty}_{1, T}$ is exponential in $n$,
\begin{equation}
  \left(\mathbb{\Lambda}^{[n+1]}_{1, T} - \mathbb{\Lambda}^{[n]}_{1, T}\right)_{ij} \propto e^{-2 \gamma (n-1)}.
\end{equation}

In what follows we present the results for the performance of an Ising quantum Otto cycle when nonadiabatic transformations are implemented.

\section{The real Ising quantum Otto cycle}
\label{sec:result_ising}

In this section we focus on a quantum Otto engine, whose working medium is constituted by
a chain of $N$ spin-$1/2$ systems interacting through the well known one dimensional
transverse-field Ising Hamiltonian:
\begin{equation}
  H^\text{spin}(t) = -J \sum_{i=1}^{N-1} \sigma_i^x \sigma_{i+1}^x - h(t) \sum_{i=1}^N \sigma_i^z,
  \label{eq:H_spin}
\end{equation}
where $\sigma^{\alpha}_i$ are the Pauli matrices ($\alpha=x,y,z$) acting on the $i$-th site.
To avoid the presence of degenerate energy eigenstates, we assume open boundary conditions.
Hereafter we also set $J=1$ as the energy scale of the system and set $h>0$ without loss of generality.
The model~\eqref{eq:H_spin} exhibits a zero-temperature quantum phase transition at $h_c = 1$,
from a paramagnetic ($h > h_c$) to a ferromagnetic ($h < h_c$) phase, after spontaneously breaking
the $\mathbb{Z}_2$ symmetry that rotates spins by a $\pi$-angle around the $z$ axis~\cite{Sachdev_book, RossiniRev_2021}.

The Hamiltonian~\eqref{eq:H_spin} can be mapped into a quadratic fermionic model
as the one in Eq.~\eqref{eq:fermionic_hamiltonian}, through a Jordan-Wigner transformation
\begin{equation}
  \sigma_i^- =  K_i c_i \, , \qquad \text{with} \quad  K_i = \Pi_{j=1}^{i-1} \sigma_j^z \, ,
\end{equation}
where $\sigma_i^\pm = ( \sigma_i^x \pm i  \sigma_i^y)/2$ are the raising/lowering operators
for the $i$-th spin. 
The mapped fermionic model is usually referred to as the Kitaev chain, which maintains the $\mathbb{Z}_2$
symmetry of the Ising chain, in the form of fermionic parity: the Ising spin-spin coupling
transforms into nearest-neighbor hopping and $p$-wave pairing terms,
while the transverse field is mapped into a chemical potential term~\cite{Kitaev_2001}.

\begin{figure}
  \centering
  \includegraphics[width=0.47\textwidth]{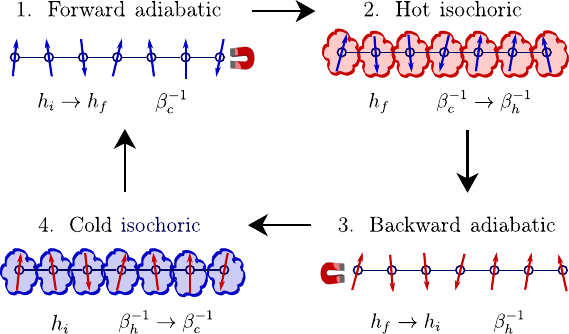}
  \caption{Sketch of the Ising quantum Otto cycle consisting of four strokes:
    (1) forward adiabatic (increasing of the transverse field),
    (2) hot isochoric (thermalization with the hot reservoir),
    (3) backward adiabatic (decreasing of the transverse field),
    (4) cold isochoric (thermalization with the cold reservoir).
    Note that, in the thermalization strokes, each spin is assumed to be coupled to
    a single bath at the same temperature $\beta_{h/c}$, in such a way that thermalization to
    that nominal temperature $\beta_{h/c}$ is guaranteed~\cite{DAbbruzzo_2021}.}
  \label{fig:sketch_cycle}
\end{figure}

One can engineer an Ising quantum Otto cycle by implementing the four transformations (strokes) discussed in
Sec.~\ref{sec:The quantum Otto cycle}, with the transverse field $h_i$ taking the role of the Hamiltonian
control parameter $\lambda$~\cite{Piccitto_2022}.
In general, for the engine being useful, the order of the transformations must be chosen carefully.
As detailed in App.~\ref{app:parametri_ising}, here we implement the following procedure,
starting from a configuration in which the working substance (the system) is supposed to be in the thermal state at
the (cold) temperature $\beta_c^{-1}$:
\begin{enumerate}
\item The system is decoupled from the external environment and the transverse field $h(t)$
  is increased linearly in a finite time $T = \delta/2$, such that $h(t) = h_i + (h_f-h_i)t/T$ (with $t \in [0,T]$),
  where $h_f > h_i$;
\item The transverse field is kept fixed at $h(t)=h_f$, while the system is put in contact with a hot reservoir at $\beta_h^{-1}$
  for a finite time $\tau = \gamma/2$ and let equilibrate with it;
\item The system is again decoupled from the external environment and the transverse field $h(t)$
  is decreased linearly in a finite time $T = \delta/2$, such that $h(t) = h_f + (h_i-h_f)t/T$ (with $t \in [0,T]$);
\item The transverse field is kept fixed at $h(t)=h_i$, while the system is put in contact with a cold reservoir at $\beta_c^{-1}$
  for a finite time $\tau = \gamma/2$ and let equilibrate with it.
\end{enumerate}
A sketch of this cycle is shown in Fig.~\ref{fig:sketch_cycle},
where the thermalization is achieved by coupling each spin to a single local bath at the same
temperature, within the Lindblad framework (see App.~\ref{app:thermalization} for details).

In Ref.~\cite{Piccitto_2022} it has been shown that, for this choice of the quench parameters with
$h_f > h_i$ (details in App.~\ref{app:parametri_ising}), in the ideal case, this engine can operate
both as a heat engine and as a refrigerator, and that the emergence of criticality in the many-body medium
may play an important role in determining the performance. 
Below we relax the assumption of ideal thermodynamic transformations (which would require an infinite operational
time and represents an unrealistic condition corresponding to zero power)
and discuss the cases where the quantum adiabaticity regime is not occurring in the strokes 1 and 3.

\subsection{Perfect thermalization}

\begin{figure}[!t]
  \includegraphics[width=0.49\textwidth]{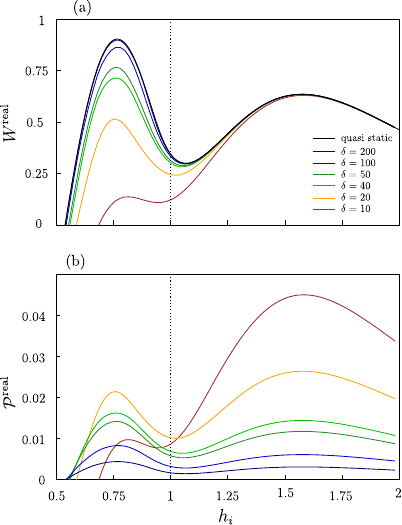}
  \caption{(a) The work performed by an imperfect Ising quantum Otto engine with $N = 50$ spins operating between
    two reservoirs at temperature $\beta_h^{-1} = 1$ and $\beta_c^{-1} = 0.5$. 
    The quench amplitude is $h_f-h_i = 0.5$.
    Different colors refer to various quench velocities. 
    We observe a more pronounced dependence on $\delta$ for quenches crossing the critical point (dotted line). 
    (b) The corresponding power output, for the same parameters of (a). 
    The plot is obtained by assuming a thermalization time $\gamma = 5$. 
    We note that in the ferromagnetic phase the power exhibits a nonmonotonic behavior in $\delta$.}
  \label{fig:work_real}
\end{figure}

Following the results of Sec.~\ref{subsec:real-adiabatic}, it is straightforward to derive the expressions
for the heat exchanged and the work performed under the assumption of perfect thermalization strokes (2 and 4),
but quantum nonadiabatic transformations:
\begin{align}
  Q^\text{real}_h = & \phantom{-} \sum_k \omega_k(h_f) \tilde{f} \big[ \beta_h, \omega_k(h_f) \big]
  \!-\! \tilde{\omega}_k(h_f) \tilde{f} \big[ \beta_c, \omega_k(h_i) \big], \nonumber \\
  Q^\text{real}_c = & - \!\sum_k \tilde{\omega}_k(h_i) \tilde{f} \big[ \beta_h, \omega_k(h_f) \big]
  \!-\! \omega_k(h_i) \tilde{f} \big[ \beta_c, \omega_k(h_i) \big], \nonumber \\
  W^\text{real}   = & \phantom{-} \sum_k \Big\{ \left[ \omega_k(h_f)- \tilde{\omega}_k(h_i) \right]
  \tilde{f} \big[ \beta_h, \omega_k(h_f) \big] \nonumber\\
  & \qquad \; - \big[ \tilde{\omega}_k(h_f) - \omega_k(h_i) \big] \tilde{f} \big[ \beta_c, \omega_k(h_i) \big] \Big\},
\end{align}
where $\tilde{f}\big[ \beta_{h(c)}, \omega_k(h_{f(i)}) \big] \equiv f \big[ \beta_{h(c)}, \omega_k(h_{f(i)}) \big] - 1/2$
and $\tilde{\omega}_k^{f(i)} = \tilde{\mathbb{H}}^{f(i)}_{kk}$, as defined in Sec.~\ref{subsec:real-adiabatic}.
Note that $W^\text{real} = Q^\text{real}_h + Q^\text{real}_c$, since the perfect thermalization strokes
reset the system state to the thermal one. We can thus define the power of the engine as
\begin{equation}
    \mathcal{P}^\text{real}  = \frac{ W^\text{real}}{\delta + \gamma}.
\end{equation}

In Fig.~\ref{fig:work_real} we show numerical results for the work (a) and the power output (b)
of an engine made of $N = 50$ spins and operating between temperatures $\beta_h^{-1} = 1$
and $\beta_c^{-1} = 0.5$, for a fixed quench amplitude $h_f-h_i = 0.5$.
Unless specified otherwise, in the following all simulations have been performed with these parameters.
Data are plotted against the initial transverse field $h_i$, while the various curves refer to different quench durations $\delta$.
To evaluate the power, we fix a thermalization time of $\gamma = 5$ so that, according to Eq.~\eqref{eq:error},
$\xi(1) \approx 5\times10^{-5}$. Under the assumption of $\mathcal{J}=1$,
this is sufficient to achieve a nearly perfect thermalization for the working medium, therefore here
the isochoric transformations can be considered ideal, although occurring in a finite time $\gamma$.

\begin{figure}
  \centering
  \includegraphics[width=0.49\textwidth]{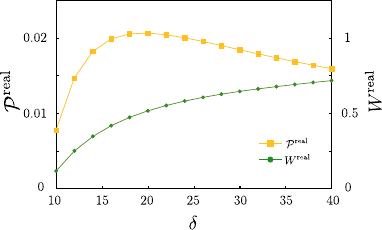}
  \caption{Power (yellow curve) and work (green curve), evaluated at $h_i = 0.76$, versus $\delta$
    (cf.~data in Fig.~\ref{fig:work_real}).
    While the work grows monotonically with the quench duration,
    the power exhibits a nonmonotonic behavior.}
  \label{fig:max_w_e_p}
\end{figure}

An important emerging feature is the sensitivity of the work on the quench duration $\delta$, for quenches occurring
in proximity to the critical point $h=1$ (marked by the dotted line), such that $h_i \lesssim 1$ and $h_f \gtrsim 1$. 
For the selected range of $\delta \in [10,200]$, this marked dependence disappears when operating
entirely in the paramagnetic phase ($h_i > 1$), due to considerably larger values of the relevant
energy gaps; in fact, a phenomenology similar to what we observe for $h$ crossing the critical point
can be recovered when considering much faster quenches ($\delta \ll 10$), becoming comparable with the spectral gap
(we checked that, for $h_i \approx 1.5$, one should take $\delta \sim 1$ to detect a clear
sensitivity in $\delta$ on the scale of the figure).

Focusing on the region $h_i \lesssim 1$ and for sufficiently large values of $\delta \gtrsim 12$, we can
identify the value $h_i \approx 0.76$ as the position of the so-called critical peak, defined as the maximum value
assumed by $W^\text{real}$ or by $\mathcal{P}^\text{real}$ in the region $h_i<1$~\cite{Piccitto_2022}.
Besides that, both the initial field $h_i$ and the quench velocity $\delta$ may modify the operation regime
of the engine ($W<0$, for $h_i$ smaller than a given threshold which depends on $\delta$):
in fact, the energy wasted in excitations could prevent the engine from producing work, i.e., the system  dissipates
so much energy that one would have to perform work on it, in order to operate the cycle.
Moreover, while curves for the work in Fig.~\ref{fig:work_real}(a) are monotonic in the quench duration,
those in Fig.~\ref{fig:work_real}(b) exhibit different scalings with $\delta$,
suggesting the existence of an optimal working point which maximizes the power delivered by the engine.

\begin{figure}[!t]
  \centering
  \includegraphics[width=0.49\textwidth]{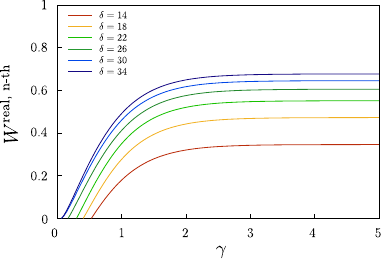}
  \caption{Work performed by the Ising Otto engine versus the thermalization time for different quench durations.
    Data are for $N=50$ spins, $h_i = 0.76$, $\beta_h^{-1} = 1$, and $\beta_c^{-1} = 0.5$.}
  \label{fig:work_vs_eta}
\end{figure}

To further investigate this feature, in Fig.~\ref{fig:max_w_e_p} we plot the value of the work (green curve)
and of the power (yellow curve) evaluated at the critical peak ($h_i = 0.76$) as a function of $\delta$.
These data display a clear nonmonotonicity of the power, contrary to the monotonic behavior of the work.
Different parameters (i.e., quench amplitudes, reservoir temperatures, and sizes of the medium)
do not qualitatively alter this scenario, although the work and power output at the optimal working point
generally depends on the size of the energy gaps that are relevant to the unitary dynamics~\cite{footnote:Paramag}.

\begin{figure*}
  \centering
  \includegraphics[width=\textwidth]{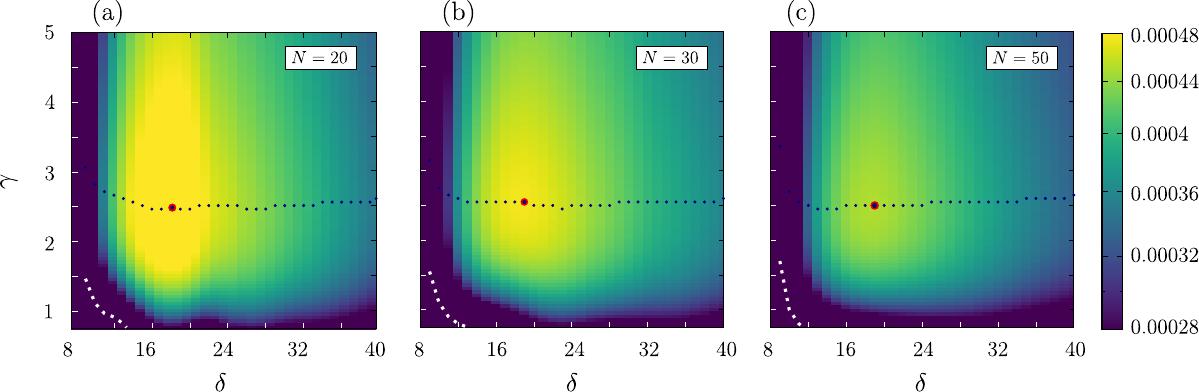}
  \caption{Contour plot of the power per spin for the nonideal quantum Ising Otto engine, as a function of $\delta$ and $\gamma$,
    for $h_i = 0.76$, $\beta_h^{-1} = 1$, and $\beta_c^{-1} = 0.5$. We adopted a logarithmic color scale,
    to better appreciate variations in the parameter space.
    The three panels refer to different lengths of the quantum Ising model (i.e., of the mapped Kitaev chain):
    $N=20$ (a), $N=30$ (b), and $N=50$ (c).
    The blue dots indicate, for any $\delta$, the value of $\gamma$ at which the maximum of the power occurs.
    The red dot locates the global maximum of the power output. Data in the bottom-left corners, below the white dots,
    are not significant, since they correspond to a parameter range where the system is not operating as a heat engine.}
  \label{fig:projected}
\end{figure*}

\subsection{Nonperfect thermalization}
\label{subsec:nontherm}

We now relax the hypothesis of perfect thermalization. In this case,
in the asymptotic regime, we find that
\begin{eqnarray}
  Q_c^\text{real, n-th} &=& \text{Tr} \big[ \, \mathbb{H}_d^f \left(\mathbb{\Lambda}_{1,T} 
 -\mathbb{\Lambda}_{2}\right) \big],
  \nonumber \\
  Q_h^\text{real, n-th} &=& \text{Tr} \big[ \mathbb{H}_d^i \left(\mathbb{\Lambda}_{2,T}  -\mathbb{\Lambda}_{1}\right) \big],
  \\
  W^\text{real, n-th} &=& Q_c + Q_h\nonumber ,
\label{eq:sistematotalmenteimperfetto}
\end{eqnarray}
where the trace is over the $2N$ degrees of freedom of the Bogoliubov matrices.

Equations~\eqref{eq:sistematotalmenteimperfetto} for the stationary heat and the work cannot be expressed analytically.
However, we can exploit Eq.~\eqref{eq:lambda_infinito} to obtain an approximate expression for the correlation matrix
that is suitable for numerical computation. In particular we can stop at order $K$, knowing we are making
an error $\varepsilon < 8N\xi(K+1)$ for work and $\varepsilon < 4N\xi(K+1)$ for heat, where $\xi(K+1)$ is defined
in Eq.~\eqref{eq:error}. This approximation breaks down when  the modulus of the work computational derived is of order $N\xi\left(K+1\right)$, in this case we cannot be sure that we are neglecting relevant terms, while for $\left|W\right| \gg 8N\xi(K+1)$ the approximation works. 

In Fig.~\ref{fig:work_vs_eta} we show the work performed by the engine at the critical peak~\cite{footnote:Initialcond},
as a function of the thermalization time $\gamma$ and for different quench durations $\delta$.
Curves are all monotonic, both in $\delta$ and in $\gamma$, and exhibit a fast (exponential) convergence
to the asymptotic value obtained for the ideal thermalization (although the thermalization time scale $\gamma$ obviously
depends on the system-bath coupling details --- see App.~\ref{app:thermalization}).
This implies a nonmonotonic behavior of the power with $\delta$ and $\gamma$, similarly to the case
of quantum nonadiabatic transformations and perfect thermalization discussed in Sec.~\ref{subsec:nontherm}. 
Note also that imperfect thermalization strokes, especially for very fast quenches, can influence the operation modes as well. 

The corresponding power per spin is plotted in Fig.~\ref{fig:projected} (color scale) as a function of $\delta$ and $\gamma$,
for various sizes of the working medium (the mapped Kitaev chain): $N = 20$ (a), $N = 30$ (b), and $N = 50$ (c). 
In each panel, the blue dots locate the position of the local maximum for fixed $\delta$, while the red dot
indicates the global one. The data in the bottom-left corner, below the white points, refer to a region
of the parameter space in which the Otto cycle is not operating as a heat engine, therefore it should not be considered.
A comparison between the three panels reveals that the system size basically affects the absolute value of the power output,
which is maximized for small $N$. This is in agreement with the fact that, at criticality, the energy gap closes 
with the system size and thus the work extraction for large $N$ is affected by fast quenches.
The functional form of the power (and, consequently, the location of the maxima) with $\delta$ and $\gamma$ seems to be independent of $N$.
Finally, the edge of the operation modes, when increasing the system size, slightly moves towards smaller $\delta$ and larger $\gamma$.

We notice that, even though some nonmonotonicity in $\gamma$ emerges, this is not as relevant as for the case discussed
in Fig.~\ref{fig:max_w_e_p}, suggesting that the role played by the thermalization time in the optimization
of the power output is marginal. 
This is not surprising when considering that the thermalization timescale is smaller than the time
needed for implementing the adiabatic sweep. 
In our analysis we assumed $\mathcal{J} = 1$; of course, a different choice 
(corresponding to an alternative modeling of the thermal reservoirs and of their coupling to
the working substance, see App.~\ref{app:thermalization}) would have lead to different thermalization timescales,
thus quantitatively modifying the analysis of the performances of the engine.

We mention that the analysis performed so far for the heat engine could be repeated for the refrigerator mode, as well.
In the perfect thermalized case, we expect the coefficient of performance (i.e., the ratio between the heat absorbed from
the cold bath $Q_c$ and the work done on the system $|W|$) to be maximized by ideal quantum adiabatic processes, as for the heat engine.
On the other hand, for $Q_c$ the analysis is more complicated: in fact, it does not depend on the velocity of the quench in which $h$ increases,
but only on the velocity of the other quench. Therefore the heat absorbed in time from the cold bath should have a monotonic behavior
with the duration of the quench in which $h$ increases (better results for sudden quenches) and a nonmonotonic behavior
with the duration of the quench in which $h$ decreases.

\section{Conclusions}
\label{sec:concl}

In this work we analyzed the performances of an imperfect quantum many-body Otto engine based on a free-fermion medium. 
After recalling the basic concepts of quantum thermodynamics and some properties of quadratic fermionic systems,
we briefly discussed the isothermal, the isochoric, and the adiabatic (unitary) transformation in this context.
We derived analytic expressions for heat and the work exchanged when such transformations are performed
under the hypotheses of perfect thermalization and quantum adiabaticity. 
Then we considered isochoric and unitary transformations in which these hypotheses of ideality are violated
(i.e., nonperfect thermalization due to a finite-time contact with an external thermal bath,
or nonperfect quantum adiabaticity due to a finite speed in the variation of the Hamiltonian control parameter).
Applying our machinery to the case of a nonperfect quantum Otto cycle, we have extended results of Ref.~\cite{Piccitto_2022},
finding that, while the work performed is always maximized in the ideal engine, the power output exhibits
a nonmonotonic behavior with the thermalization time $\tau= \gamma/2$ and the quench time $T=\delta/2$,
suggesting the existence of an optimal working point in the $(\tau, T)$ parameter space. 

A possible extension of this research is directed towards a systematic characterization of quantum heat engines
based on the spectral properties of the Hamiltonian. As an example, it could be interesting to predict analytically
the operational mode and the performances expected, as well as to derive analytic conditions to identify
the optimal working points, once the Hamiltonian details are provided. One could also study quenches
with a time pattern different than linear~\cite{Sen_2008, Barankov_2008} (e.g., in such a way that the variation
of the Hamiltonian parameters are faster far from criticality and slower near to it) or
more complicated approaches favoring adiabatic dynamics~\cite{Muga_2019}:
this kind of schemes may help to increase the power of the engine and thus change its optimal working point.

As a future perspective, it would be also tempting to generalize our results to working media composed of truly interacting
quantum many-body systems: this may represent an important step to understand the role of interactions in the work extraction process. 
Moreover, even though here we mainly focused on the average values of work and power, additional important information
can be obtained by characterizing the fluctuations of the observed quantities (both for the ideal and for the real scenario),
in order to determine whether these kind of engines can be practically useful
for the work production~\cite{Richens_2016, Holubec2017, Denzler2020, Lutz2021}. 
Finally, one could investigate alternative strategies to enhance the engine performance by considering
a working medium coupled to nonMarkovian~\cite{Wiedmann_2020} or even nonthermal baths~\cite{Scully_2003, Rossnagel_2014}
(e.g., some reservoir suitably engineered to invert the population of the system).

We hope that our analysis will serve as a guidance to any possible experimental realizations of many-body quantum engines,
that are far from being ideal. Proposals to experimentally realize the Kitaev chain with arrays
of coupled superconducting quantum dots have been recently put forward~\cite{DasSarma_2012, Kouwenhoven_2023},
indicating a possible strategy to build up such kind of devices.

\acknowledgments

We acknowledge useful discussions with M. Campisi and G. Chiriac{\`o}.
GP acknowledges financial support from PNRR MUR Project PE0000023-NQSTI.

\appendix

\section{Thermalization of free-fermion system}
\label{app:thermalization}
  
To perform the two isochoric transformations, we suppose that the working medium is in contact with
an environment made of $N$ identical thermal baths, described by a continuum of fermionic harmonic oscillators
that ensures proper thermalization in the asymptotic long coupling-time limit.
The environment Hamiltonian reads
\begin{equation}
	H_\text{env}  = \sum_i \int dq \,\varepsilon_i(q) \, \kappa^\dagger_i(q)\, \kappa_i(q),
\end{equation}
with $\varepsilon_i(q) \ge 0$ and $\kappa_i^{(\dagger)}(q)$ being the annihilation (creation) fermionic operators. 
We assume the baths to be independent, therefore the reduced density matrix of the environment
can be written in the factorized form 
\begin{equation}
  \rho_\text{env} = \bigotimes_i \rho_\text{$i$-th bath},
\end{equation}
where $\rho_\text{$i$-th bath}$ is the thermal density matrix describing the $i$-th bath at temperature $\beta^{-1}$
(that corresponds to $\beta_c^{-1}$ and $\beta_h^{-1}$ for the thermalization with the cold and hot reservoir, respectively).

We assume each of the above baths to be coupled to a single site of the chain through a quadratic, factorizable, Hamiltonian
\begin{equation}
  H_\text{int} = \sum_i \int dq \, g_i(q) \, \big (c_i + c_i^\dagger \big) \, \big[ \kappa_i(q) + \kappa^\dagger_i(q) \big].
\end{equation}
Then we introduce the density of states of the baths 
\begin{equation}
  \mathcal{J}_i(\omega) \equiv \pi \int dq \, |g_i(q)|^2 \, \delta \big[ \omega-\varepsilon_i(k) \big] \,.
\end{equation}
Assuming a very large bath bandwidth with respect to the system frequencies, one has
$\mathcal{J}_i(\omega) \simeq \mathcal{J}_i$.
By tracing out the environmental degrees of freedom and imposing the Born-Markov approximation for the baths,
we derive, in the energy eigenbasis, the microscopic Lindblad master equation for the reduced density matrix
of the system, whose stationary solution is a thermal state at temperature $T$~\cite{Petruccione_book, DAbbruzzo_2021}.
From this equation and under the hypothesis of no degeneracies in the spectrum,
we can analytically derive the expressions for the Bogoliubov quasiparticles correlator at the time $\tau$
of the thermalization process:
\begin{align}
  \braket{b_k^\dagger b_k}(\tau) = & \, f(\beta, \omega_k) \big( 1 - e^{-2\mathcal{J} \tau} \big)
  + \braket{b_k^\dagger b_k}(0) \, e^{-2\mathcal{J}\tau}, \nonumber \\
  \braket{b_k^\dagger b_q}(\tau) = & \braket{b^{\dagger}_k b_q}_{\rho_0} e^{i(\omega_k - \omega_q)\tau -2 \mathcal{J}t}, \nonumber \\
  \braket{b^{\dagger}_k b^{\dagger}_q}(\tau) = & \braket{b^{\dagger}_k b^{\dagger}_q}_{\rho_0} e^{i (\omega_k + \omega_q)\tau  -2 \mathcal{J}t}, \nonumber\\
   \braket{b_k b_q}(\tau) = & \braket{b_k b_q}_{\rho_0} e^{-i(\omega_k + \omega_q)\tau  -2 \mathcal{J}t}.  
\end{align}
Note that we assumed the same density of state $\mathcal{J}_i = \mathcal{J}$ for all the local baths.
The first expression coincides with Eq.~\eqref{eq:correlator_nonth} with $\gamma = 2\tau$,
while the other expressions are the ones reported in Eqs.~\eqref{eq:Bath_corr} (in the latter, $\mathcal{J}=1$).

\section{Thermodynamic limit of the Ideal Ising Otto engine}
\label{App:ising_tdlimit}

In the thermodynamic limit $N \to + \infty$, it is possible to obtain a simple analytical expression
for the eigenvalues of $\mathbb{H}^\text{spin}$:
\begin{equation}
\omega_k(h) = 2\sqrt{1 + h^2 - 2h \cos(k)}, ~ k \in (0, \pi].
	\label{eq:ising_spectrum}
\end{equation}
In this limit, even though the adiabatic theorem is not valid, because of the critical closure of the gap
for $h=1$, it is instructive to study the behavior of the heat and the work per spin, defined as
\begin{equation}
q_{h(c)} = \lim_{N \to + \infty} Q_{h(c)}/N, \hspace{0.3cm} w = \lim_{N \to + \infty} W/N.
\end{equation}
By substituting $\sum_{k=1}^{N} \mapsto \frac{N}{\pi} \int_{0}^{\pi}dk$ in Eqs.~\eqref{eq:heat}, we find
\begin{subequations}
\begin{eqnarray}
  q_h^\text{id} & = & \phantom{-} \frac{2}{\pi} \int_{0}^{\pi}{dk \hspace{0.1cm} \omega_k(h_f) \, \big( \Delta f_k \big)_{hc}}\,, \\
  q_c^\text{id} & = &          -  \frac{2}{\pi} \int_{0}^{\pi}{dk \hspace{0.1cm} \omega_k(h_i) \, \big( \Delta f_k \big)_{hc}}\,,\\
  w^\text{id}   & = & \phantom{-} \frac{2}{\pi} \int_{0}^{\pi}{dk \hspace{0.1cm} \big[ \omega_k(h_f) - \omega_k(h_i) \big] \big( \Delta f_k \big)_{hc}}\,,\qquad
\end{eqnarray}
\end{subequations}
where $w^\text{id} = q^\text{id}_h + q^\text{id}_c$, 
while the ratio in Eq.~\eqref{eq:ratio} becomes
\begin{equation}
    r_k\left(h_i,h_f\right) = \sqrt{\frac{1+h^2_i-2h_i\cos(k)}{1+h^2_f-2h_f\cos(k)}} \,.
\end{equation}

\begin{figure}
  \includegraphics[width=0.4\textwidth]{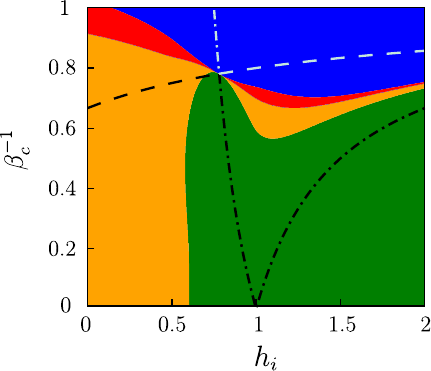}
  \caption{Operating modes of the ideal Ising Otto engine in the thermodynamic limit ($N \to \infty$)
    and for fixed $\beta^{-1}_h = 1$, $h_f - h_i = 0.5$. The color code indicates the operating mode:
    orange stands for a thermal accelerator, green for a heat engine, red for a heater and blue for a refrigerator.
    All the four modes are possible, as shown numerically for finite $N$, in Ref.~\cite{Piccitto_2022}.
    The dashed and dashed-dotted lines represent, respectively, the equations $|1- h_i| / |1- h_f| = \beta_h / \beta_c$
    and $(1 + h_i)/(1 + h_f) = \beta_h / \beta_c$. The intersection point is the Carnot point.
    The area in between the white portions of such lines is a parameter subspace in which the system can only operate
    as a refrigerator; the area under the black portions of such lines is where the system can operate either
    as a heat engine or a heat accelerator.}
  \label{fig:operatingmodes}
\end{figure}

Now, taking the derivative of the square of the ratio,
\begin{equation}
  \frac{\partial r_{k}^2(h_i,h_f)}{\partial k} = \frac{2\left(h_f h_i - 1 \right)\left(h_f-h_i\right)\sin(k)}
       {\big[ 1 + h_f^{2} - 2h_f\cos(k) \big]^{2}} 
\end{equation}
and using the fact that $h_i < h_f$ for construction, we obtain that,
if $h_i > h_f^{-1}$ then $\partial_{k} r_{k}^2(h_i,h_f) > 0$, otherwise $\partial_{k} r_{k}^2(h_i,h_f) < 0$.
So, if $h_i > h_f^{-1}$, then $r_{k}(h_i,h_f)$ is an increasing function for $k \in (0,\pi]$
and plugging the dispersion relation in \ref{item:dispersionrelation} we have:
\begin{itemize}
   \item for $r_{k \to 0^+}(h_i,h_f) \equiv \frac{|1- h_i|}{|1- h_f|}> \beta_h / \beta_c$ the system is a heat engine;\\
   \item for $r_{k=\pi}(h_i,h_f)    \equiv \frac{1+ h_i}{1+h_f} < \beta_h / \beta_c$ the system is a refrigerator.
\end{itemize}
On the other hand, if $h_i < h_f^{-1}$, then $r_{\kappa}(h_i,h_f)$ is a decreasing function for $k \in (0,\pi]$, thus we have:
\begin{itemize}
   \item for $r_{k=\pi}(h_i,h_f)    \equiv \frac{1+h_i}{1+h_f} > \beta_h / \beta_c$ the system could be a heat engine or a thermal accelerator;\\
   \item for $r_{k \to 0^+}(h_i,h_f) \equiv \frac{|1-h_i|}{|1-h_f|} < \beta_h / \beta_c$ the system is a refrigerator.
\end{itemize}
The remaining case $h_i = h_f^{-1}$ is a special point, since in that case $r_k$ is constant ($r_k = h_i$).
This allows us to say that, for $\beta_h / \beta_c < h_i < 1$, the system is a heat engine,
while for $h_i < \beta_h / \beta_c$ it performs as a refrigerator.
The point $h_i = h_f^{-1} = \beta_h / \beta_c$ is called a Carnot point,
since in that case no work is produced nor the heat is exchanged, and the efficiency of the engine saturates the Carnot bound.

Figure~\ref{fig:operatingmodes} shows a contour plot of the modes in which the Ising Otto engine in the thermodynamic limit
can operate, as a function of the initial field strength $h_i$ and of the cold reservoir temperature $\beta^{-1}_c$,
while we fix $\beta^{-1}_h = 1$ and $h_f - h_i = 0.5$ (see caption).
This complements the numerical analysis performed in Ref.~\cite{Piccitto_2022} for finite values of $N$,
since it is obtained with a fully analytic approach and by directly accessing the thermodynamic limit.

\section{Work in the Ising Otto engine}
\label{app:parametri_ising}

In this section we comment on the order of the adiabatic strokes that are necessary to obtain an Otto engine
that can produce useful work. For classical engines, in which the work is univocally associated to a variation
of mechanical energy, the situation is rather intuitive: the cold working medium is adiabatically compressed and
then warmed up in order to further increase the energy that, during the adiabatic expansion, is released, thus producing work. 
This order can be quite easily extended to the case of few-body quantum engines (e.g., a collection of noninteracting qubits,
as in Ref.~\cite{Solfanelli_2020}), in which the adiabatic compression (expansion) corresponds to a compression (expansion)
of the spectrum of the system.

The situation becomes more involved when considering many-body quantum systems, since is not immediate to find
  a way to stretch or compress the full eigenspectrum.
We thus call a compression (an expansion) any quench that increases (decreases) the internal energy of the system.
In quantum heat engines the work can be associated
to a variation of some physical quantity that is not necessarily the mechanical energy. 
As a consequence, one does not need to keep the order of the transformations unaltered, and, on the contrary,
it can also be reversed.
  
\begin{figure}
  \centering
  \includegraphics[width=0.49\textwidth]{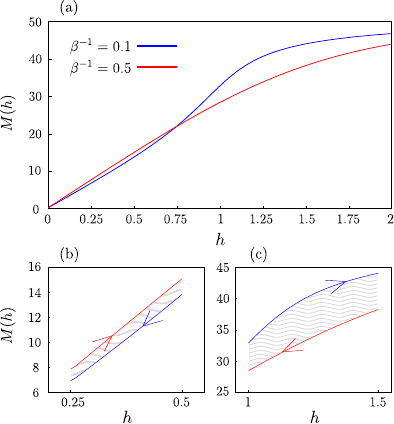}
  \caption{(a): Behavior of $M(h)$ in function of the transverse field $h$ for $\beta^{-1} = 0.1$ (blue curve) and $\beta^{-1}=0.5$ (red curve).
    (b): Inset of panel (a) for quench from $h_i = 0.25$ to $h_f=0.5$. The arrows represent in which direction the transverse field must be varied at any temperature to perform, by implementing sequentially the two transformations, the total work represented by the shaded area.
    (c): The same of panel (b) for quench from $h_i = 1$ to $h_f=1.5$.}
  \label{fig:fgen}
\end{figure}

As an example, let us consider the Ising quantum engine. 
In this case, the work is associated to a change of the magnetization $M(h)$ (see Refs.~\cite{Dorner_2012, Fusco_2014}).
Hence, the work performed by the Ising chain during a quasistatic isothermal transformation at a fixed temperature $\beta^{-1}$
can be evaluated according to Eq.~\eqref{eq:work_qs}:
\begin{equation}
  W = \int_{h_i}^{h_f} dh \, M(h), \qquad \text{with}\  M(h) \equiv -\frac{\partial \mathcal{F}}{\partial h}.
\end{equation}
In Fig.~\ref{fig:fgen}(a) we show the function $M(h)$ for two different temperatures $\beta^{-1} = 0.1$ (blue curve)
and $\beta^{-1} = 0.5$ (red curve) versus the transverse field.
If $h$ is varied from smaller to larger values, the work performed during the transformation is the area
beneath such curve. If the transformation is performed backward, the work acquires a minus sign, compared as before.

We now consider a sequence of four transformations implementing a Carnot cycle: a forward quasistatic variation of the transverse
field while the system is in thermal equilibrium at $\beta^{-1}_1$, a thermalization with the bath at temperature $\beta^{-1}_2$,
a backward quasistatic variation of $h$ while the system is in thermal equilibrium at $\beta^{-1}_2$,
and a thermalization with the bath at temperature $\beta_1^{-1}$.
The total work of such an engine is the sum of the works performed along the two adiabatic strokes.
Therefore, the temperatures $\beta_1^{-1}$ and $\beta_2^{-1}$ must be chosen in order to follow the curves clockwise. 
As an example, if the Hamiltonian parameter is varied from $h_i = 0.25$ to $h_f = 0.5$, as in Fig.~\ref{fig:fgen}(b),
the only possible choice is $\beta^{-1}_1 = 0.5$ and $\beta_2^{-1} = 0.1$.
Conversely, considering a quench from $h_i=1$ to $h_f=1.5$, as in Fig.~\ref{fig:fgen}(c),
one should choose $\beta_1^{-1} = 0.1$ and $\beta_2^{-1} = 0.5$. 
In both figures, the shaded area represents the work performed by the engine.

For a generic fermionic quadratic system as the one in Eq.~\eqref{eq:fermionic_hamiltonian},
the order to perform the cycle can be argued by looking at the spectrum.
If $\omega'_k(\lambda) \geq 0 $, $\forall \lambda \in [\lambda_i,\lambda_f]$, $\forall k$,
(here $\lambda$ denotes the control parameter)
the work output is positive (produced by the system) for increasing $\lambda$, since,
in the interval $\lambda \in [\lambda_i,\lambda_f]$,
\begin{equation}
   -\frac{\partial \mathcal{F}}{\partial \lambda} =  \sum_k \omega'_k(\lambda)
  \Big[ \tfrac12- f\big[\beta,\omega_k(\lambda)\big] \Big] > 0.
\end{equation}
Beside this, we have that
\begin{equation}
    -\frac{\partial}{\partial \beta} \frac{\partial \mathcal{F}}{\partial \lambda} = \sum_k \omega'_k(\lambda)
  \Big[ \omega_k(\lambda) e^{\beta \omega_k(\lambda)}f^2\big[\beta,\omega_k(\lambda)\big]\Big]>0,  
  \label{eq:derivatabeta}
\end{equation}
where we used the fact that $\omega_k(\lambda) > 0$ $\forall k$.
Equation~\eqref{eq:derivatabeta} proves that the work produced is larger at small temperatures.
As a consequence, the order of the cycle should follow that in Fig.~\ref{fig:fgen}(c), in which the forward
adiabatic transformation is performed at $\beta_c$.
If, on the other hand, $\omega'_k(\lambda) \leq 0 $, $\forall \lambda \in [\lambda_i,\lambda_f]$, $\forall k$,
using the same arguments as in the latter case, we conclude that, to produce work,
the order of the transformations must be reversed.

Even though, to simplify the discussion, here we have only considered quasistatic transformations
(instead of the adiabatic processes for the Otto cycle, we adopted the isotherm processes for the Carnot cycle),
we can use the above results to gain some insights for the adiabatic transformation of the quantum Otto Ising cycle, as well.
In particular, since $W^\text{ad} \le W^\text{iso}$, a necessary condition for the Otto cycle to operate as a heat engine
is that the Carnot cycle operating between the same temperatures ($\beta_c^{-1}$, $\beta_h^{-1}$) produces work.
For example, in Fig.~\ref{fig:work_real}, to obtain an operative heat engine for quenches of the Ising chain
from $h_i \in [0.5, 2]$ to $h_f=h_i+0.5$, we have chosen as the initial temperature $\beta_1^{-1} \equiv \beta_c^{-1} = 0.5$.

\bibliography{biblio_Otto}

\begin{thebibliography}{65}%
\makeatletter
\providecommand \@ifxundefined [1]{%
 \@ifx{#1\undefined}
}%
\providecommand \@ifnum [1]{%
 \ifnum #1\expandafter \@firstoftwo
 \else \expandafter \@secondoftwo
 \fi
}%
\providecommand \@ifx [1]{%
 \ifx #1\expandafter \@firstoftwo
 \else \expandafter \@secondoftwo
 \fi
}%
\providecommand \natexlab [1]{#1}%
\providecommand \enquote  [1]{``#1''}%
\providecommand \bibnamefont  [1]{#1}%
\providecommand \bibfnamefont [1]{#1}%
\providecommand \citenamefont [1]{#1}%
\providecommand \href@noop [0]{\@secondoftwo}%
\providecommand \href [0]{\begingroup \@sanitize@url \@href}%
\providecommand \@href[1]{\@@startlink{#1}\@@href}%
\providecommand \@@href[1]{\endgroup#1\@@endlink}%
\providecommand \@sanitize@url [0]{\catcode `\\12\catcode `\$12\catcode
  `\&12\catcode `\#12\catcode `\^12\catcode `\_12\catcode `\%12\relax}%
\providecommand \@@startlink[1]{}%
\providecommand \@@endlink[0]{}%
\providecommand \url  [0]{\begingroup\@sanitize@url \@url }%
\providecommand \@url [1]{\endgroup\@href {#1}{\urlprefix }}%
\providecommand \urlprefix  [0]{URL }%
\providecommand \Eprint [0]{\href }%
\providecommand \doibase [0]{https://doi.org/}%
\providecommand \selectlanguage [0]{\@gobble}%
\providecommand \bibinfo  [0]{\@secondoftwo}%
\providecommand \bibfield  [0]{\@secondoftwo}%
\providecommand \translation [1]{[#1]}%
\providecommand \BibitemOpen [0]{}%
\providecommand \bibitemStop [0]{}%
\providecommand \bibitemNoStop [0]{.\EOS\space}%
\providecommand \EOS [0]{\spacefactor3000\relax}%
\providecommand \BibitemShut  [1]{\csname bibitem#1\endcsname}%
\let\auto@bib@innerbib\@empty
\bibitem [{\citenamefont {Alicki}(1979)}]{Alicki_1979}%
  \BibitemOpen
  \bibfield  {author} {\bibinfo {author} {\bibfnamefont {R.}~\bibnamefont
  {Alicki}},\ }\bibfield  {title} {\bibinfo {title} {{The quantum open system
  as a model of the heat engine}},\ }\href
  {https://doi.org/10.1088/0305-4470/12/5/007} {\bibfield  {journal} {\bibinfo
  {journal} {J. Phys. A: Math. Gen.}\ }\textbf {\bibinfo {volume} {12}},\
  \bibinfo {pages} {L103} (\bibinfo {year} {1979})}\BibitemShut {NoStop}%
\bibitem [{\citenamefont {Kosloff}(1984)}]{Kosloff1984}%
  \BibitemOpen
  \bibfield  {author} {\bibinfo {author} {\bibfnamefont {R.}~\bibnamefont
  {Kosloff}},\ }\bibfield  {title} {\bibinfo {title} {{A quantum mechanical
  open system as a model of a heat engine}},\ }\href
  {https://doi.org/10.1063/1.446862} {\bibfield  {journal} {\bibinfo  {journal}
  {J. Chem. Phys.}\ }\textbf {\bibinfo {volume} {80}},\ \bibinfo {pages} {1625}
  (\bibinfo {year} {1984})}\BibitemShut {NoStop}%
\bibitem [{\citenamefont {Scovil}\ and\ \citenamefont
  {Schulz-DuBois}(1959)}]{Scovi_1959}%
  \BibitemOpen
  \bibfield  {author} {\bibinfo {author} {\bibfnamefont {H.~E.}\ \bibnamefont
  {Scovil}}\ and\ \bibinfo {author} {\bibfnamefont {E.~O.}\ \bibnamefont
  {Schulz-DuBois}},\ }\bibfield  {title} {\bibinfo {title} {Three-level masers
  as heat engines},\ }\href {https://doi.org/10.1103/PhysRevLett.2.262}
  {\bibfield  {journal} {\bibinfo  {journal} {Phys. Rev. Lett.}\ }\textbf
  {\bibinfo {volume} {2}},\ \bibinfo {pages} {262} (\bibinfo {year}
  {1959})}\BibitemShut {NoStop}%
\bibitem [{\citenamefont {Quan}\ \emph {et~al.}(2007)\citenamefont {Quan},
  \citenamefont {Liu}, \citenamefont {Sun},\ and\ \citenamefont
  {Nori}}]{Quan_2007}%
  \BibitemOpen
  \bibfield  {author} {\bibinfo {author} {\bibfnamefont {H.~T.}\ \bibnamefont
  {Quan}}, \bibinfo {author} {\bibfnamefont {Y.-x.}\ \bibnamefont {Liu}},
  \bibinfo {author} {\bibfnamefont {C.~P.}\ \bibnamefont {Sun}},\ and\ \bibinfo
  {author} {\bibfnamefont {F.}~\bibnamefont {Nori}},\ }\bibfield  {title}
  {\bibinfo {title} {Quantum thermodynamic cycles and quantum heat engines},\
  }\href {https://doi.org/10.1103/PhysRevE.76.031105} {\bibfield  {journal}
  {\bibinfo  {journal} {Phys. Rev. E}\ }\textbf {\bibinfo {volume} {76}},\
  \bibinfo {pages} {031105} (\bibinfo {year} {2007})}\BibitemShut {NoStop}%
\bibitem [{\citenamefont {Rossnagel}\ \emph {et~al.}(2014)\citenamefont
  {Rossnagel}, \citenamefont {Abah}, \citenamefont {Schmidt-Kaler},
  \citenamefont {Singer},\ and\ \citenamefont {Lutz}}]{Rossnagel_2014}%
  \BibitemOpen
  \bibfield  {author} {\bibinfo {author} {\bibfnamefont {J.}~\bibnamefont
  {Rossnagel}}, \bibinfo {author} {\bibfnamefont {O.}~\bibnamefont {Abah}},
  \bibinfo {author} {\bibfnamefont {F.}~\bibnamefont {Schmidt-Kaler}}, \bibinfo
  {author} {\bibfnamefont {K.}~\bibnamefont {Singer}},\ and\ \bibinfo {author}
  {\bibfnamefont {E.}~\bibnamefont {Lutz}},\ }\bibfield  {title} {\bibinfo
  {title} {Nanoscale heat engine beyond the carnot limit},\ }\href
  {https://doi.org/10.1103/PhysRevLett.112.030602} {\bibfield  {journal}
  {\bibinfo  {journal} {Phys. Rev. Lett.}\ }\textbf {\bibinfo {volume} {112}},\
  \bibinfo {pages} {030602} (\bibinfo {year} {2014})}\BibitemShut {NoStop}%
\bibitem [{\citenamefont {Hartmann}\ \emph
  {et~al.}(2020{\natexlab{a}})\citenamefont {Hartmann}, \citenamefont
  {Mukherjee}, \citenamefont {Niedenzu},\ and\ \citenamefont
  {Lechner}}]{hartmann2020A}%
  \BibitemOpen
  \bibfield  {author} {\bibinfo {author} {\bibfnamefont {A.}~\bibnamefont
  {Hartmann}}, \bibinfo {author} {\bibfnamefont {V.}~\bibnamefont {Mukherjee}},
  \bibinfo {author} {\bibfnamefont {W.}~\bibnamefont {Niedenzu}},\ and\
  \bibinfo {author} {\bibfnamefont {W.}~\bibnamefont {Lechner}},\ }\bibfield
  {title} {\bibinfo {title} {{Many-body quantum heat engines with shortcuts to
  adiabaticity}},\ }\href {https://doi.org/10.1103/PhysRevResearch.2.023145}
  {\bibfield  {journal} {\bibinfo  {journal} {Phys. Rev. Res.}\ }\textbf
  {\bibinfo {volume} {2}},\ \bibinfo {pages} {023145} (\bibinfo {year}
  {2020}{\natexlab{a}})}\BibitemShut {NoStop}%
\bibitem [{\citenamefont {Kosloff}\ and\ \citenamefont
  {Rezek}(2017)}]{Kosloff-rev2017}%
  \BibitemOpen
  \bibfield  {author} {\bibinfo {author} {\bibfnamefont {R.}~\bibnamefont
  {Kosloff}}\ and\ \bibinfo {author} {\bibfnamefont {Y.}~\bibnamefont
  {Rezek}},\ }\bibfield  {title} {\bibinfo {title} {{The quantum harmonic Otto
  cycle}},\ }\href {https://doi.org/10.3390/e19040136} {\bibfield  {journal}
  {\bibinfo  {journal} {Entropy}\ }\textbf {\bibinfo {volume} {19}},\ \bibinfo
  {pages} {136} (\bibinfo {year} {2017})}\BibitemShut {NoStop}%
\bibitem [{\citenamefont {Bender}\ \emph {et~al.}(2000)\citenamefont {Bender},
  \citenamefont {Brody},\ and\ \citenamefont {Meister}}]{Bender2000}%
  \BibitemOpen
  \bibfield  {author} {\bibinfo {author} {\bibfnamefont {C.~M.}\ \bibnamefont
  {Bender}}, \bibinfo {author} {\bibfnamefont {D.~C.}\ \bibnamefont {Brody}},\
  and\ \bibinfo {author} {\bibfnamefont {B.~K.}\ \bibnamefont {Meister}},\
  }\bibfield  {title} {\bibinfo {title} {{Quantum mechanical Carnot engine}},\
  }\href {https://doi.org/10.1088/0305-4470/33/24/302} {\bibfield  {journal}
  {\bibinfo  {journal} {J. Phys. A: Math. Gen.}\ }\textbf {\bibinfo {volume}
  {33}},\ \bibinfo {pages} {4427} (\bibinfo {year} {2000})}\BibitemShut
  {NoStop}%
\bibitem [{\citenamefont {Bender}\ \emph {et~al.}(2002)\citenamefont {Bender},
  \citenamefont {Brody},\ and\ \citenamefont {Meister}}]{bender2002entropy}%
  \BibitemOpen
  \bibfield  {author} {\bibinfo {author} {\bibfnamefont {C.~M.}\ \bibnamefont
  {Bender}}, \bibinfo {author} {\bibfnamefont {D.~C.}\ \bibnamefont {Brody}},\
  and\ \bibinfo {author} {\bibfnamefont {B.~K.}\ \bibnamefont {Meister}},\
  }\bibfield  {title} {\bibinfo {title} {Entropy and temperature of a quantum
  carnot engine},\ }\href {https://doi.org/10.1098/rspa.2001.0928} {\bibfield
  {journal} {\bibinfo  {journal} {Proc. R. Soc. A: Math. Phys. Eng. Sci.}\
  }\textbf {\bibinfo {volume} {458}},\ \bibinfo {pages} {1519} (\bibinfo {year}
  {2002})}\BibitemShut {NoStop}%
\bibitem [{\citenamefont {Solfanelli}\ \emph {et~al.}(2020)\citenamefont
  {Solfanelli}, \citenamefont {Falsetti},\ and\ \citenamefont
  {Campisi}}]{Solfanelli_2020}%
  \BibitemOpen
  \bibfield  {author} {\bibinfo {author} {\bibfnamefont {A.}~\bibnamefont
  {Solfanelli}}, \bibinfo {author} {\bibfnamefont {M.}~\bibnamefont
  {Falsetti}},\ and\ \bibinfo {author} {\bibfnamefont {M.}~\bibnamefont
  {Campisi}},\ }\bibfield  {title} {\bibinfo {title} {{Nonadiabatic
  single-qubit quantum Otto engine}},\ }\href
  {https://doi.org/10.1103/PhysRevB.101.054513} {\bibfield  {journal} {\bibinfo
   {journal} {Phys. Rev. B}\ }\textbf {\bibinfo {volume} {101}},\ \bibinfo
  {pages} {054513} (\bibinfo {year} {2020})}\BibitemShut {NoStop}%
\bibitem [{\citenamefont {Henrich}\ \emph {et~al.}(2007)\citenamefont
  {Henrich}, \citenamefont {Rempp},\ and\ \citenamefont
  {Mahler}}]{Henrich2007}%
  \BibitemOpen
  \bibfield  {author} {\bibinfo {author} {\bibfnamefont {M.~J.}\ \bibnamefont
  {Henrich}}, \bibinfo {author} {\bibfnamefont {F.}~\bibnamefont {Rempp}},\
  and\ \bibinfo {author} {\bibfnamefont {G.}~\bibnamefont {Mahler}},\
  }\bibfield  {title} {\bibinfo {title} {{Quantum thermodynamic Otto machines:
  A spin-system approach}},\ }\href
  {https://doi.org/10.1140/epjst/e2007-00371-8} {\bibfield  {journal} {\bibinfo
   {journal} {Eur. Phys. J.: Spec. Top.}\ }\textbf {\bibinfo {volume} {151}},\
  \bibinfo {pages} {157} (\bibinfo {year} {2007})}\BibitemShut {NoStop}%
\bibitem [{\citenamefont {Uzdin}\ and\ \citenamefont
  {Kosloff}(2014)}]{Uzdin2014}%
  \BibitemOpen
  \bibfield  {author} {\bibinfo {author} {\bibfnamefont {R.}~\bibnamefont
  {Uzdin}}\ and\ \bibinfo {author} {\bibfnamefont {R.}~\bibnamefont
  {Kosloff}},\ }\bibfield  {title} {\bibinfo {title} {{Universal features in
  the efficiency at maximal work of hot quantum Otto engines}},\ }\href
  {https://doi.org/10.1209/0295-5075/108/40001} {\bibfield  {journal} {\bibinfo
   {journal} {Eur. Phys. Lett.}\ }\textbf {\bibinfo {volume} {108}},\ \bibinfo
  {pages} {40001} (\bibinfo {year} {2014})}\BibitemShut {NoStop}%
\bibitem [{\citenamefont {Leggio}\ and\ \citenamefont
  {Antezza}(2016)}]{Leggio2016}%
  \BibitemOpen
  \bibfield  {author} {\bibinfo {author} {\bibfnamefont {B.}~\bibnamefont
  {Leggio}}\ and\ \bibinfo {author} {\bibfnamefont {M.}~\bibnamefont
  {Antezza}},\ }\bibfield  {title} {\bibinfo {title} {{Otto engine beyond its
  standard quantum limit}},\ }\href
  {https://doi.org/10.1103/PhysRevE.93.022122} {\bibfield  {journal} {\bibinfo
  {journal} {Phys. Rev. E}\ }\textbf {\bibinfo {volume} {93}},\ \bibinfo
  {pages} {022122} (\bibinfo {year} {2016})}\BibitemShut {NoStop}%
\bibitem [{\citenamefont {Rossnagel}\ \emph {et~al.}(2016)\citenamefont
  {Rossnagel}, \citenamefont {Dawkins}, \citenamefont {Tolazzi}, \citenamefont
  {Abah}, \citenamefont {Lutz}, \citenamefont {{Schmidt-Kaler}},\ and\
  \citenamefont {Singer}}]{Rossnagel_2016}%
  \BibitemOpen
  \bibfield  {author} {\bibinfo {author} {\bibfnamefont {J.}~\bibnamefont
  {Rossnagel}}, \bibinfo {author} {\bibfnamefont {S.}~\bibnamefont {Dawkins}},
  \bibinfo {author} {\bibfnamefont {K.}~\bibnamefont {Tolazzi}}, \bibinfo
  {author} {\bibfnamefont {O.}~\bibnamefont {Abah}}, \bibinfo {author}
  {\bibfnamefont {E.}~\bibnamefont {Lutz}}, \bibinfo {author} {\bibfnamefont
  {F.}~\bibnamefont {{Schmidt-Kaler}}},\ and\ \bibinfo {author} {\bibfnamefont
  {K.}~\bibnamefont {Singer}},\ }\bibfield  {title} {\bibinfo {title} {{A
  single-atom heat engine}},\ }\href {https://doi.org/10.1126/science.aad6320}
  {\bibfield  {journal} {\bibinfo  {journal} {Science}\ }\textbf {\bibinfo
  {volume} {352}},\ \bibinfo {pages} {325} (\bibinfo {year}
  {2016})}\BibitemShut {NoStop}%
\bibitem [{\citenamefont {Maslennikov}\ \emph {et~al.}(2019)\citenamefont
  {Maslennikov}, \citenamefont {Ding}, \citenamefont {Habl{\"u}tzel},
  \citenamefont {Gan}, \citenamefont {Roulet}, \citenamefont {Nimmrichter},
  \citenamefont {Dai}, \citenamefont {Scarani},\ and\ \citenamefont
  {Matsukevich}}]{TrappedIons2019}%
  \BibitemOpen
  \bibfield  {author} {\bibinfo {author} {\bibfnamefont {G.}~\bibnamefont
  {Maslennikov}}, \bibinfo {author} {\bibfnamefont {S.}~\bibnamefont {Ding}},
  \bibinfo {author} {\bibfnamefont {R.}~\bibnamefont {Habl{\"u}tzel}}, \bibinfo
  {author} {\bibfnamefont {J.}~\bibnamefont {Gan}}, \bibinfo {author}
  {\bibfnamefont {A.}~\bibnamefont {Roulet}}, \bibinfo {author} {\bibfnamefont
  {S.}~\bibnamefont {Nimmrichter}}, \bibinfo {author} {\bibfnamefont
  {J.}~\bibnamefont {Dai}}, \bibinfo {author} {\bibfnamefont {V.}~\bibnamefont
  {Scarani}},\ and\ \bibinfo {author} {\bibfnamefont {D.}~\bibnamefont
  {Matsukevich}},\ }\bibfield  {title} {\bibinfo {title} {{Quantum absorption
  refrigerator with trapped ions}},\ }\href
  {https://doi.org/10.1038/s41467-018-08090-0} {\bibfield  {journal} {\bibinfo
  {journal} {Nat. Commun.}\ }\textbf {\bibinfo {volume} {10}},\ \bibinfo
  {pages} {202} (\bibinfo {year} {2019})}\BibitemShut {NoStop}%
\bibitem [{\citenamefont {{von Lindenfels}}\ \emph {et~al.}(2019)\citenamefont
  {{von Lindenfels}}, \citenamefont {Gr{\"a}b}, \citenamefont {Schmiegelow},
  \citenamefont {Kaushal}, \citenamefont {Schulz}, \citenamefont {Mitchison},
  \citenamefont {Goold}, \citenamefont {Schmidt-Kaler},\ and\ \citenamefont
  {Poschinger}}]{vonLindenfels2019}%
  \BibitemOpen
  \bibfield  {author} {\bibinfo {author} {\bibfnamefont {D.}~\bibnamefont {{von
  Lindenfels}}}, \bibinfo {author} {\bibfnamefont {O.}~\bibnamefont
  {Gr{\"a}b}}, \bibinfo {author} {\bibfnamefont {C.~T.}\ \bibnamefont
  {Schmiegelow}}, \bibinfo {author} {\bibfnamefont {V.}~\bibnamefont
  {Kaushal}}, \bibinfo {author} {\bibfnamefont {J.}~\bibnamefont {Schulz}},
  \bibinfo {author} {\bibfnamefont {M.~T.}\ \bibnamefont {Mitchison}}, \bibinfo
  {author} {\bibfnamefont {J.}~\bibnamefont {Goold}}, \bibinfo {author}
  {\bibfnamefont {F.}~\bibnamefont {Schmidt-Kaler}},\ and\ \bibinfo {author}
  {\bibfnamefont {U.~G.}\ \bibnamefont {Poschinger}},\ }\bibfield  {title}
  {\bibinfo {title} {{Spin heat engine coupled to a harmonic-oscillator
  flywheel}},\ }\href {https://doi.org/10.1103/PhysRevLett.123.080602}
  {\bibfield  {journal} {\bibinfo  {journal} {Phys. Rev. Lett.}\ }\textbf
  {\bibinfo {volume} {123}},\ \bibinfo {pages} {080602} (\bibinfo {year}
  {2019})}\BibitemShut {NoStop}%
\bibitem [{\citenamefont {Peterson}\ \emph {et~al.}(2019)\citenamefont
  {Peterson}, \citenamefont {Batalh{\~n a}o}, \citenamefont {Herrera},
  \citenamefont {Souza}, \citenamefont {Sarthour}, \citenamefont {Oliveira},\
  and\ \citenamefont {Serra}}]{Peterson2019}%
  \BibitemOpen
  \bibfield  {author} {\bibinfo {author} {\bibfnamefont {J.~P.}\ \bibnamefont
  {Peterson}}, \bibinfo {author} {\bibfnamefont {T.~B.}\ \bibnamefont
  {Batalh{\~n a}o}}, \bibinfo {author} {\bibfnamefont {M.}~\bibnamefont
  {Herrera}}, \bibinfo {author} {\bibfnamefont {A.~M.}\ \bibnamefont {Souza}},
  \bibinfo {author} {\bibfnamefont {R.~S.}\ \bibnamefont {Sarthour}}, \bibinfo
  {author} {\bibfnamefont {I.~S.}\ \bibnamefont {Oliveira}},\ and\ \bibinfo
  {author} {\bibfnamefont {R.~M.}\ \bibnamefont {Serra}},\ }\bibfield  {title}
  {\bibinfo {title} {{Experimental characterization of a spin quantum heat
  engine}},\ }\href {https://doi.org/10.1103/PhysRevLett.123.240601} {\bibfield
   {journal} {\bibinfo  {journal} {Phys. Rev. Lett.}\ }\textbf {\bibinfo
  {volume} {123}},\ \bibinfo {pages} {240601} (\bibinfo {year}
  {2019})}\BibitemShut {NoStop}%
\bibitem [{\citenamefont {Sheng}\ \emph {et~al.}(2021)\citenamefont {Sheng},
  \citenamefont {Yang},\ and\ \citenamefont {Wu}}]{Sheng2021}%
  \BibitemOpen
  \bibfield  {author} {\bibinfo {author} {\bibfnamefont {J.}~\bibnamefont
  {Sheng}}, \bibinfo {author} {\bibfnamefont {C.}~\bibnamefont {Yang}},\ and\
  \bibinfo {author} {\bibfnamefont {H.}~\bibnamefont {Wu}},\ }\bibfield
  {title} {\bibinfo {title} {{Realization of a coupled-mode heat engine with
  cavity-mediated nanoresonators}},\ }\href
  {https://doi.org/10.1126/sciadv.abl7740} {\bibfield  {journal} {\bibinfo
  {journal} {Sci. Adv.}\ }\textbf {\bibinfo {volume} {7}},\ \bibinfo {pages}
  {eabl7740} (\bibinfo {year} {2021})}\BibitemShut {NoStop}%
\bibitem [{\citenamefont {Bouton}\ \emph {et~al.}(2021)\citenamefont {Bouton},
  \citenamefont {Nettersheim}, \citenamefont {Burgardt}, \citenamefont {Adam},
  \citenamefont {Lutz},\ and\ \citenamefont {Widera}}]{Bouton2021}%
  \BibitemOpen
  \bibfield  {author} {\bibinfo {author} {\bibfnamefont {Q.}~\bibnamefont
  {Bouton}}, \bibinfo {author} {\bibfnamefont {J.}~\bibnamefont {Nettersheim}},
  \bibinfo {author} {\bibfnamefont {S.}~\bibnamefont {Burgardt}}, \bibinfo
  {author} {\bibfnamefont {D.}~\bibnamefont {Adam}}, \bibinfo {author}
  {\bibfnamefont {E.}~\bibnamefont {Lutz}},\ and\ \bibinfo {author}
  {\bibfnamefont {A.}~\bibnamefont {Widera}},\ }\bibfield  {title} {\bibinfo
  {title} {{A quantum heat engine driven by atomic collisions}},\ }\href
  {https://doi.org/10.1038/s41467-021-22222-z} {\bibfield  {journal} {\bibinfo
  {journal} {Nat. Commun.}\ }\textbf {\bibinfo {volume} {12}},\ \bibinfo
  {pages} {2063} (\bibinfo {year} {2021})}\BibitemShut {NoStop}%
\bibitem [{\citenamefont {Myers}\ \emph {et~al.}(2022)\citenamefont {Myers},
  \citenamefont {Abah},\ and\ \citenamefont {Deffner}}]{Myers22AQS4}%
  \BibitemOpen
  \bibfield  {author} {\bibinfo {author} {\bibfnamefont {N.~M.}\ \bibnamefont
  {Myers}}, \bibinfo {author} {\bibfnamefont {O.}~\bibnamefont {Abah}},\ and\
  \bibinfo {author} {\bibfnamefont {S.}~\bibnamefont {Deffner}},\ }\bibfield
  {title} {\bibinfo {title} {Quantum thermodynamic devices: From theoretical
  proposals to experimental reality},\ }\href
  {https://doi.org/10.1116/5.0083192} {\bibfield  {journal} {\bibinfo
  {journal} {AVS Quantum Science}\ }\textbf {\bibinfo {volume} {4}},\ \bibinfo
  {pages} {027101} (\bibinfo {year} {2022})}\BibitemShut {NoStop}%
\bibitem [{\citenamefont {Mukherjee}\ and\ \citenamefont
  {Divakaran}(2021)}]{Mukherjee2021}%
  \BibitemOpen
  \bibfield  {author} {\bibinfo {author} {\bibfnamefont {V.}~\bibnamefont
  {Mukherjee}}\ and\ \bibinfo {author} {\bibfnamefont {U.}~\bibnamefont
  {Divakaran}},\ }\bibfield  {title} {\bibinfo {title} {{Many-body quantum
  thermal machines}},\ }\href {https://doi.org/10.1088/1361-648X/ac1b60}
  {\bibfield  {journal} {\bibinfo  {journal} {J. Phys.: Condens. Matter}\
  }\textbf {\bibinfo {volume} {33}},\ \bibinfo {pages} {454001} (\bibinfo
  {year} {2021})}\BibitemShut {NoStop}%
\bibitem [{\citenamefont {Cangemi}\ \emph {et~al.}(2023)\citenamefont
  {Cangemi}, \citenamefont {Bhadra},\ and\ \citenamefont {Levy}}]{Cangemi2023}%
  \BibitemOpen
  \bibfield  {author} {\bibinfo {author} {\bibfnamefont {L.~M.}\ \bibnamefont
  {Cangemi}}, \bibinfo {author} {\bibfnamefont {C.}~\bibnamefont {Bhadra}},\
  and\ \bibinfo {author} {\bibfnamefont {A.}~\bibnamefont {Levy}},\ }\href
  {https://doi.org/10.48550/arXiv.2302.00726} {} (\bibinfo {year} {2023}),\
  \Eprint {https://arxiv.org/abs/2302.00726} {arXiv:2302.00726 [quant-ph]}
  \BibitemShut {NoStop}%
\bibitem [{\citenamefont {Jaramillo}\ \emph {et~al.}(2016)\citenamefont
  {Jaramillo}, \citenamefont {Beau},\ and\ \citenamefont {del
  Campo}}]{jaramillo2016quantum}%
  \BibitemOpen
  \bibfield  {author} {\bibinfo {author} {\bibfnamefont {J.}~\bibnamefont
  {Jaramillo}}, \bibinfo {author} {\bibfnamefont {M.}~\bibnamefont {Beau}},\
  and\ \bibinfo {author} {\bibfnamefont {A.}~\bibnamefont {del Campo}},\
  }\bibfield  {title} {\bibinfo {title} {Quantum supremacy of many-particle
  thermal machines},\ }\href
  {https://doi.org/10.1088%2F1367-2630%2F18%2F7%2F075019} {\bibfield  {journal}
  {\bibinfo  {journal} {New J. Phys.}\ }\textbf {\bibinfo {volume} {18}},\
  \bibinfo {pages} {075019} (\bibinfo {year} {2016})}\BibitemShut {NoStop}%
\bibitem [{\citenamefont {Bengtsson}\ \emph {et~al.}(2018)\citenamefont
  {Bengtsson}, \citenamefont {Tengstrand}, \citenamefont {Wacker},
  \citenamefont {Samuelsson}, \citenamefont {Ueda}, \citenamefont {Linke},\
  and\ \citenamefont {Reimann}}]{Bengtsson2018}%
  \BibitemOpen
  \bibfield  {author} {\bibinfo {author} {\bibfnamefont {J.}~\bibnamefont
  {Bengtsson}}, \bibinfo {author} {\bibfnamefont {M.~N.}\ \bibnamefont
  {Tengstrand}}, \bibinfo {author} {\bibfnamefont {A.}~\bibnamefont {Wacker}},
  \bibinfo {author} {\bibfnamefont {P.}~\bibnamefont {Samuelsson}}, \bibinfo
  {author} {\bibfnamefont {M.}~\bibnamefont {Ueda}}, \bibinfo {author}
  {\bibfnamefont {H.}~\bibnamefont {Linke}},\ and\ \bibinfo {author}
  {\bibfnamefont {S.~M.}\ \bibnamefont {Reimann}},\ }\bibfield  {title}
  {\bibinfo {title} {{Quantum Szilard engine with attractively interacting
  bosons}},\ }\href {https://doi.org/10.1103/PhysRevLett.120.100601} {\bibfield
   {journal} {\bibinfo  {journal} {Phys. Rev. Lett.}\ }\textbf {\bibinfo
  {volume} {120}},\ \bibinfo {pages} {100601} (\bibinfo {year}
  {2018})}\BibitemShut {NoStop}%
\bibitem [{\citenamefont {Chen}\ \emph {et~al.}(2019)\citenamefont {Chen},
  \citenamefont {Watanabe}, \citenamefont {Yu}, \citenamefont {Guan},\ and\
  \citenamefont {{del Campo}}}]{Chen2019}%
  \BibitemOpen
  \bibfield  {author} {\bibinfo {author} {\bibfnamefont {Y.-Y.}\ \bibnamefont
  {Chen}}, \bibinfo {author} {\bibfnamefont {G.}~\bibnamefont {Watanabe}},
  \bibinfo {author} {\bibfnamefont {Y.-C.}\ \bibnamefont {Yu}}, \bibinfo
  {author} {\bibfnamefont {X.-W.}\ \bibnamefont {Guan}},\ and\ \bibinfo
  {author} {\bibfnamefont {A.}~\bibnamefont {{del Campo}}},\ }\bibfield
  {title} {\bibinfo {title} {{An interaction-driven many-particle quantum heat
  engine and its universal behavior}},\ }\href
  {https://doi.org/10.1038/s41534-019-0204-5} {\bibfield  {journal} {\bibinfo
  {journal} {npj Quantum Inf.}\ }\textbf {\bibinfo {volume} {5}},\ \bibinfo
  {pages} {88} (\bibinfo {year} {2019})}\BibitemShut {NoStop}%
\bibitem [{\citenamefont {Fogarty}\ and\ \citenamefont
  {Busch}(2020)}]{Fogarty2020}%
  \BibitemOpen
  \bibfield  {author} {\bibinfo {author} {\bibfnamefont {T.}~\bibnamefont
  {Fogarty}}\ and\ \bibinfo {author} {\bibfnamefont {T.}~\bibnamefont
  {Busch}},\ }\bibfield  {title} {\bibinfo {title} {{A many-body heat engine at
  criticality}},\ }\href {https://doi.org/10.1088/2058-9565/abbc63} {\bibfield
  {journal} {\bibinfo  {journal} {Quantum Sci. Technol.}\ }\textbf {\bibinfo
  {volume} {6}},\ \bibinfo {pages} {015003} (\bibinfo {year}
  {2020})}\BibitemShut {NoStop}%
\bibitem [{\citenamefont {Carollo}\ \emph {et~al.}(2020)\citenamefont
  {Carollo}, \citenamefont {Gambetta}, \citenamefont {Brandner}, \citenamefont
  {Garrahan},\ and\ \citenamefont {Lesanovsky}}]{Carollo2020}%
  \BibitemOpen
  \bibfield  {author} {\bibinfo {author} {\bibfnamefont {F.}~\bibnamefont
  {Carollo}}, \bibinfo {author} {\bibfnamefont {F.~M.}\ \bibnamefont
  {Gambetta}}, \bibinfo {author} {\bibfnamefont {K.}~\bibnamefont {Brandner}},
  \bibinfo {author} {\bibfnamefont {J.~P.}\ \bibnamefont {Garrahan}},\ and\
  \bibinfo {author} {\bibfnamefont {I.}~\bibnamefont {Lesanovsky}},\ }\bibfield
   {title} {\bibinfo {title} {{Nonequilibrium quantum many-body Rydberg atom
  engine}},\ }\href {https://doi.org/10.1103/PhysRevLett.124.170602} {\bibfield
   {journal} {\bibinfo  {journal} {Phys. Rev. Lett.}\ }\textbf {\bibinfo
  {volume} {124}},\ \bibinfo {pages} {170602} (\bibinfo {year}
  {2020})}\BibitemShut {NoStop}%
\bibitem [{\citenamefont {Boubakour}\ \emph {et~al.}(2023)\citenamefont
  {Boubakour}, \citenamefont {Fogarty},\ and\ \citenamefont
  {Busch}}]{Boubakour2023}%
  \BibitemOpen
  \bibfield  {author} {\bibinfo {author} {\bibfnamefont {M.}~\bibnamefont
  {Boubakour}}, \bibinfo {author} {\bibfnamefont {T.}~\bibnamefont {Fogarty}},\
  and\ \bibinfo {author} {\bibfnamefont {T.}~\bibnamefont {Busch}},\ }\bibfield
   {title} {\bibinfo {title} {{Interaction-enhanced quantum heat engine}},\
  }\href {https://doi.org/10.1103/PhysRevResearch.5.013088} {\bibfield
  {journal} {\bibinfo  {journal} {Phys. Rev. Res.}\ }\textbf {\bibinfo {volume}
  {5}},\ \bibinfo {pages} {013088} (\bibinfo {year} {2023})}\BibitemShut
  {NoStop}%
\bibitem [{\citenamefont {Watson}\ and\ \citenamefont
  {Kheruntsyan}(2023)}]{Watson2023}%
  \BibitemOpen
  \bibfield  {author} {\bibinfo {author} {\bibfnamefont {R.~S.}\ \bibnamefont
  {Watson}}\ and\ \bibinfo {author} {\bibfnamefont {K.~V.}\ \bibnamefont
  {Kheruntsyan}},\ }\href {https://arxiv.org/abs/2308.05266} {} (\bibinfo
  {year} {2023}),\ \Eprint {https://arxiv.org/abs/2308.05266} {arXiv:2308.05266
  [cond-mat]} \BibitemShut {NoStop}%
\bibitem [{\citenamefont {Halpern}\ \emph {et~al.}(2019)\citenamefont
  {Halpern}, \citenamefont {White}, \citenamefont {Gopalakrishnan},\ and\
  \citenamefont {Refael}}]{Halpern2019}%
  \BibitemOpen
  \bibfield  {author} {\bibinfo {author} {\bibfnamefont {N.~Y.}\ \bibnamefont
  {Halpern}}, \bibinfo {author} {\bibfnamefont {C.~D.}\ \bibnamefont {White}},
  \bibinfo {author} {\bibfnamefont {S.}~\bibnamefont {Gopalakrishnan}},\ and\
  \bibinfo {author} {\bibfnamefont {G.}~\bibnamefont {Refael}},\ }\bibfield
  {title} {\bibinfo {title} {{Quantum engine based on many-body
  localization}},\ }\href {https://doi.org/10.1103/PhysRevB.99.024203}
  {\bibfield  {journal} {\bibinfo  {journal} {Phys. Rev. B}\ }\textbf {\bibinfo
  {volume} {99}},\ \bibinfo {pages} {024203} (\bibinfo {year}
  {2019})}\BibitemShut {NoStop}%
\bibitem [{\citenamefont {Hartmann}\ \emph
  {et~al.}(2020{\natexlab{b}})\citenamefont {Hartmann}, \citenamefont
  {Mukherjee}, \citenamefont {Mbeng}, \citenamefont {Niedenzu},\ and\
  \citenamefont {Lechner}}]{hartmann2020B}%
  \BibitemOpen
  \bibfield  {author} {\bibinfo {author} {\bibfnamefont {A.}~\bibnamefont
  {Hartmann}}, \bibinfo {author} {\bibfnamefont {V.}~\bibnamefont {Mukherjee}},
  \bibinfo {author} {\bibfnamefont {G.~B.}\ \bibnamefont {Mbeng}}, \bibinfo
  {author} {\bibfnamefont {W.}~\bibnamefont {Niedenzu}},\ and\ \bibinfo
  {author} {\bibfnamefont {W.}~\bibnamefont {Lechner}},\ }\bibfield  {title}
  {\bibinfo {title} {{Multi-spin counter-diabatic driving in many-body quantum
  Otto refrigerators}},\ }\href {https://doi.org/10.22331/q-2020-12-24-377}
  {\bibfield  {journal} {\bibinfo  {journal} {Quantum}\ }\textbf {\bibinfo
  {volume} {4}},\ \bibinfo {pages} {377} (\bibinfo {year}
  {2020}{\natexlab{b}})}\BibitemShut {NoStop}%
\bibitem [{\citenamefont {Wang}(2138)}]{Wang2020}%
  \BibitemOpen
  \bibfield  {author} {\bibinfo {author} {\bibfnamefont {Q.}~\bibnamefont
  {Wang}},\ }\bibfield  {title} {\bibinfo {title} {{Performance of quantum heat
  engines under the influence of long-range interactions}},\ }\href
  {https://doi.org/10.1103/PhysRevE.102.012138} {\bibfield  {journal} {\bibinfo
   {journal} {Phys. Rev. E}\ }\textbf {\bibinfo {volume} {102}},\ \bibinfo
  {pages} {012138} (\bibinfo {year} {012138})}\BibitemShut {NoStop}%
\bibitem [{\citenamefont {Revathy}\ \emph {et~al.}(2020)\citenamefont
  {Revathy}, \citenamefont {Mukherjee}, \citenamefont {Divakaran},\ and\
  \citenamefont {{del Campo}}}]{delcampo2020}%
  \BibitemOpen
  \bibfield  {author} {\bibinfo {author} {\bibfnamefont {B.~S.}\ \bibnamefont
  {Revathy}}, \bibinfo {author} {\bibfnamefont {V.}~\bibnamefont {Mukherjee}},
  \bibinfo {author} {\bibfnamefont {U.}~\bibnamefont {Divakaran}},\ and\
  \bibinfo {author} {\bibfnamefont {A.}~\bibnamefont {{del Campo}}},\
  }\bibfield  {title} {\bibinfo {title} {{Universal finite-time thermodynamics
  of many-body quantum machines from Kibble-Zurek scaling}},\ }\href
  {https://doi.org/10.1103/PhysRevResearch.2.043247} {\bibfield  {journal}
  {\bibinfo  {journal} {Phys. Rev. Res.}\ }\textbf {\bibinfo {volume} {2}},\
  \bibinfo {pages} {043247} (\bibinfo {year} {2020})}\BibitemShut {NoStop}%
\bibitem [{\citenamefont {Piccitto}\ \emph {et~al.}(2022)\citenamefont
  {Piccitto}, \citenamefont {Campisi},\ and\ \citenamefont
  {Rossini}}]{Piccitto_2022}%
  \BibitemOpen
  \bibfield  {author} {\bibinfo {author} {\bibfnamefont {G.}~\bibnamefont
  {Piccitto}}, \bibinfo {author} {\bibfnamefont {M.}~\bibnamefont {Campisi}},\
  and\ \bibinfo {author} {\bibfnamefont {D.}~\bibnamefont {Rossini}},\
  }\bibfield  {title} {\bibinfo {title} {The ising critical quantum otto
  engine},\ }\href {https://doi.org/10.1088/1367-2630/ac963b} {\bibfield
  {journal} {\bibinfo  {journal} {New J. Phys.}\ }\textbf {\bibinfo {volume}
  {24}},\ \bibinfo {pages} {103023} (\bibinfo {year} {2022})}\BibitemShut
  {NoStop}%
\bibitem [{\citenamefont {Solfanelli}\ \emph {et~al.}(2023)\citenamefont
  {Solfanelli}, \citenamefont {Giachetti}, \citenamefont {Campisi},
  \citenamefont {Ruffo},\ and\ \citenamefont {Defenu}}]{Solfanelli_2023}%
  \BibitemOpen
  \bibfield  {author} {\bibinfo {author} {\bibfnamefont {A.}~\bibnamefont
  {Solfanelli}}, \bibinfo {author} {\bibfnamefont {G.}~\bibnamefont
  {Giachetti}}, \bibinfo {author} {\bibfnamefont {M.}~\bibnamefont {Campisi}},
  \bibinfo {author} {\bibfnamefont {S.}~\bibnamefont {Ruffo}},\ and\ \bibinfo
  {author} {\bibfnamefont {N.}~\bibnamefont {Defenu}},\ }\bibfield  {title}
  {\bibinfo {title} {Quantum heat engine with long-range advantages},\ }\href
  {https://doi.org/10.1088/1367-2630/acc04e} {\bibfield  {journal} {\bibinfo
  {journal} {New J. Phys.}\ }\textbf {\bibinfo {volume} {25}},\ \bibinfo
  {pages} {033030} (\bibinfo {year} {2023})}\BibitemShut {NoStop}%
\bibitem [{\citenamefont {Williamson}\ and\ \citenamefont
  {Davis}(2023)}]{Williamson2024}%
  \BibitemOpen
  \bibfield  {author} {\bibinfo {author} {\bibfnamefont {L.~A.}\ \bibnamefont
  {Williamson}}\ and\ \bibinfo {author} {\bibfnamefont {M.~J.}\ \bibnamefont
  {Davis}},\ }\bibfield  {title} {\bibinfo {title} {{Many-body enhancement in a
  spin-chain quantum heat engine}},\ }\href
  {https://doi.org/10.1103/PhysRevB.109.024310} {\bibfield  {journal} {\bibinfo
   {journal} {Phys. Rev. B}\ }\textbf {\bibinfo {volume} {109}},\ \bibinfo
  {pages} {024310} (\bibinfo {year} {2023})}\BibitemShut {NoStop}%
\bibitem [{\citenamefont {Campisi}\ and\ \citenamefont
  {Fazio}(2016)}]{Campisi_2016}%
  \BibitemOpen
  \bibfield  {author} {\bibinfo {author} {\bibfnamefont {M.}~\bibnamefont
  {Campisi}}\ and\ \bibinfo {author} {\bibfnamefont {R.}~\bibnamefont
  {Fazio}},\ }\bibfield  {title} {\bibinfo {title} {{The power of a critical
  heat engine}},\ }\href {https://doi.org/10.1038/ncomms11895} {\bibfield
  {journal} {\bibinfo  {journal} {Nat. Commun.}\ }\textbf {\bibinfo {volume}
  {7}},\ \bibinfo {pages} {11895} (\bibinfo {year} {2016})}\BibitemShut
  {NoStop}%
\bibitem [{\citenamefont {{Del Grosso}}\ \emph {et~al.}(2022)\citenamefont
  {{Del Grosso}}, \citenamefont {Lombardo}, \citenamefont {Mazzitelli},\ and\
  \citenamefont {Villar}}]{DelGrosso2022}%
  \BibitemOpen
  \bibfield  {author} {\bibinfo {author} {\bibfnamefont {N.~F.}\ \bibnamefont
  {{Del Grosso}}}, \bibinfo {author} {\bibfnamefont {F.~C.}\ \bibnamefont
  {Lombardo}}, \bibinfo {author} {\bibfnamefont {F.~D.}\ \bibnamefont
  {Mazzitelli}},\ and\ \bibinfo {author} {\bibfnamefont {P.~I.}\ \bibnamefont
  {Villar}},\ }\bibfield  {title} {\bibinfo {title} {{Quantum Otto cycle in a
  superconducting cavity in the nonadiabatic regime}},\ }\href
  {https://doi.org/10.1103/PhysRevA.105.022202} {\bibfield  {journal} {\bibinfo
   {journal} {Phys. Rev. A}\ }\textbf {\bibinfo {volume} {105}},\ \bibinfo
  {pages} {022202} (\bibinfo {year} {2022})}\BibitemShut {NoStop}%
\bibitem [{foo({\natexlab{a}})}]{footnote:ksum}%
  \BibitemOpen
  \href@noop {} {\bibinfo {title} {{Below, when not specified, we will always
  assume summation indexes running from $1$ to $N$.}}}
  ({\natexlab{a}})\BibitemShut {NoStop}%
\bibitem [{\citenamefont {Sakurai}\ and\ \citenamefont
  {Napolitano}(2020)}]{Sakurai_book}%
  \BibitemOpen
  \bibfield  {author} {\bibinfo {author} {\bibfnamefont {J.~J.}\ \bibnamefont
  {Sakurai}}\ and\ \bibinfo {author} {\bibfnamefont {J.}~\bibnamefont
  {Napolitano}},\ }\href {https://doi.org/10.1017/9781108587280} {\emph
  {\bibinfo {title} {{Modern Quantum Mechanics, 3rd edition}}}}\ (\bibinfo
  {publisher} {Cambridge University Press},\ \bibinfo {year}
  {2020})\BibitemShut {NoStop}%
\bibitem [{foo({\natexlab{b}})}]{footnote:Adiab}%
  \BibitemOpen
  \href@noop {} {\bibinfo {title} {{We remind that a transformation can be
  considered approximately adiabatic if the variation of the system is carried
  on a time $\delta \gg |\lambda_i - \lambda_f| / \Delta^2$, where $\Delta$
  quantifies the relevant energy scale of the model.}}}
  ({\natexlab{b}})\BibitemShut {NoStop}%
\bibitem [{\citenamefont {{D'Abbruzzo}}\ and\ \citenamefont
  {Rossini}(2021)}]{DAbbruzzo_2021}%
  \BibitemOpen
  \bibfield  {author} {\bibinfo {author} {\bibfnamefont {A.}~\bibnamefont
  {{D'Abbruzzo}}}\ and\ \bibinfo {author} {\bibfnamefont {D.}~\bibnamefont
  {Rossini}},\ }\bibfield  {title} {\bibinfo {title} {{Self-consistent
  microscopic derivation of Markovian master equations for open quadratic
  quantum systems}},\ }\href {https://doi.org/10.1103/PhysRevA.103.052209}
  {\bibfield  {journal} {\bibinfo  {journal} {Phys. Rev. A}\ }\textbf {\bibinfo
  {volume} {103}},\ \bibinfo {pages} {052209} (\bibinfo {year}
  {2021})}\BibitemShut {NoStop}%
\bibitem [{foo({\natexlab{c}})}]{footnote:Diag}%
  \BibitemOpen
  \href@noop {} {\bibinfo {title} {{These matrices are diagonal, hence it is
  possible to work with a single vector containing the diagonal elements of
  $\mathbb{\Theta}_{c(h)}$ and $\mathbb{\Gamma}^{[n]}_{c(h)}$. However, we
  prefer to adopt the matrix formalism to simplify the generalization of these
  calculations in the case of a cycle with real adiabatic transformations.}}}
  ({\natexlab{c}})\BibitemShut {NoStop}%
\bibitem [{\citenamefont {Mbeng}\ \emph {et~al.}(2020)\citenamefont {Mbeng},
  \citenamefont {Russomanno},\ and\ \citenamefont {Santoro}}]{Mbeng_2020}%
  \BibitemOpen
  \bibfield  {author} {\bibinfo {author} {\bibfnamefont {G.~B.}\ \bibnamefont
  {Mbeng}}, \bibinfo {author} {\bibfnamefont {A.}~\bibnamefont {Russomanno}},\
  and\ \bibinfo {author} {\bibfnamefont {G.~E.}\ \bibnamefont {Santoro}},\
  }\href {https://doi.org/10.48550/arXiv.2009.09208} {} (\bibinfo {year}
  {2020}),\ \Eprint {https://arxiv.org/abs/2009.09208} {arXiv:2009.09208
  [quant-ph]} \BibitemShut {NoStop}%
\bibitem [{\citenamefont {Piccitto}\ \emph {et~al.}(2023)\citenamefont
  {Piccitto}, \citenamefont {Russomanno},\ and\ \citenamefont
  {Rossini}}]{Piccitto_2023}%
  \BibitemOpen
  \bibfield  {author} {\bibinfo {author} {\bibfnamefont {G.}~\bibnamefont
  {Piccitto}}, \bibinfo {author} {\bibfnamefont {A.}~\bibnamefont
  {Russomanno}},\ and\ \bibinfo {author} {\bibfnamefont {D.}~\bibnamefont
  {Rossini}},\ }\bibfield  {title} {\bibinfo {title} {Entanglement dynamics
  with string measurement operators},\ }\href
  {https://doi.org/10.21468/scipostphyscore.6.4.078} {\bibfield  {journal}
  {\bibinfo  {journal} {SciPost Phys. Core}\ }\textbf {\bibinfo {volume} {6}},\
  \bibinfo {pages} {078} (\bibinfo {year} {2023})}\BibitemShut {NoStop}%
\bibitem [{\citenamefont {Allahverdyan}\ \emph {et~al.}(2004)\citenamefont
  {Allahverdyan}, \citenamefont {Balian},\ and\ \citenamefont
  {Nieuwenhuizen}}]{allahverdyan2004maximal}%
  \BibitemOpen
  \bibfield  {author} {\bibinfo {author} {\bibfnamefont {A.~E.}\ \bibnamefont
  {Allahverdyan}}, \bibinfo {author} {\bibfnamefont {R.}~\bibnamefont
  {Balian}},\ and\ \bibinfo {author} {\bibfnamefont {T.~M.}\ \bibnamefont
  {Nieuwenhuizen}},\ }\bibfield  {title} {\bibinfo {title} {Maximal work
  extraction from finite quantum systems},\ }\href
  {https://doi.org/10.1209/epl/i2004-10101-2} {\bibfield  {journal} {\bibinfo
  {journal} {Europhys. Lett.}\ }\textbf {\bibinfo {volume} {67}},\ \bibinfo
  {pages} {565} (\bibinfo {year} {2004})}\BibitemShut {NoStop}%
\bibitem [{\citenamefont {Sachdev}(2011)}]{Sachdev_book}%
  \BibitemOpen
  \bibfield  {author} {\bibinfo {author} {\bibfnamefont {S.}~\bibnamefont
  {Sachdev}},\ }\href {https://doi.org/10.1017/CBO9780511973765} {\emph
  {\bibinfo {title} {{Quantum Phase Transitions, 2nd edition}}}}\ (\bibinfo
  {publisher} {Cambridge University Press},\ \bibinfo {year}
  {2011})\BibitemShut {NoStop}%
\bibitem [{\citenamefont {Rossini}\ and\ \citenamefont
  {Vicari}(2021)}]{RossiniRev_2021}%
  \BibitemOpen
  \bibfield  {author} {\bibinfo {author} {\bibfnamefont {D.}~\bibnamefont
  {Rossini}}\ and\ \bibinfo {author} {\bibfnamefont {E.}~\bibnamefont
  {Vicari}},\ }\bibfield  {title} {\bibinfo {title} {Coherent and dissipative
  dynamics at quantum phase transitions},\ }\href
  {https://doi.org/10.1016/j.physrep.2021.08.003} {\bibfield  {journal}
  {\bibinfo  {journal} {Phys. Rep.}\ }\textbf {\bibinfo {volume} {936}},\
  \bibinfo {pages} {1–110} (\bibinfo {year} {2021})}\BibitemShut {NoStop}%
\bibitem [{\citenamefont {Kitaev}(2001)}]{Kitaev_2001}%
  \BibitemOpen
  \bibfield  {author} {\bibinfo {author} {\bibfnamefont {A.~Y.}\ \bibnamefont
  {Kitaev}},\ }\bibfield  {title} {\bibinfo {title} {{Unpaired Majorana
  fermions in quantum wires}},\ }\href
  {https://doi.org/10.1070/1063-7869/44/10s/s29} {\bibfield  {journal}
  {\bibinfo  {journal} {Phys.-Usp.}\ }\textbf {\bibinfo {volume} {44}},\
  \bibinfo {pages} {131–136} (\bibinfo {year} {2001})}\BibitemShut {NoStop}%
\bibitem [{foo({\natexlab{d}})}]{footnote:Paramag}%
  \BibitemOpen
  \href@noop {} {\bibinfo {title} {{In the paramegnetic region ($h_i>1$) and
  for the same set of parameters used in {Fig}.~\ref{fig:max_w_e_p}, one is
  however able to achieve a significantly larger power output, since smaller
  values of $\delta$ are sufficient for the engine to deliver a comparable
  work.}}} ({\natexlab{d}})\BibitemShut {NoStop}%
\bibitem [{foo({\natexlab{e}})}]{footnote:Initialcond}%
  \BibitemOpen
  \href@noop {} {\bibinfo {title} {{We checked that the qualitative behavior of
  the work and power output is quite insensitive to the specific value of
  $h_i<1$, therefore we are not showing data for other initial conditions.}}}
  ({\natexlab{e}})\BibitemShut {NoStop}%
\bibitem [{\citenamefont {Sen}\ \emph {et~al.}(2008)\citenamefont {Sen},
  \citenamefont {Sengupta},\ and\ \citenamefont {Mondal}}]{Sen_2008}%
  \BibitemOpen
  \bibfield  {author} {\bibinfo {author} {\bibfnamefont {D.}~\bibnamefont
  {Sen}}, \bibinfo {author} {\bibfnamefont {K.}~\bibnamefont {Sengupta}},\ and\
  \bibinfo {author} {\bibfnamefont {S.}~\bibnamefont {Mondal}},\ }\bibfield
  {title} {\bibinfo {title} {{Defect production in nonlinear quench across a
  quantum critical point}},\ }\href
  {https://doi.org/10.1103/PhysRevLett.101.016806} {\bibfield  {journal}
  {\bibinfo  {journal} {Phys. Rev. Lett.}\ }\textbf {\bibinfo {volume} {101}},\
  \bibinfo {pages} {016806} (\bibinfo {year} {2008})}\BibitemShut {NoStop}%
\bibitem [{\citenamefont {Barankov}\ and\ \citenamefont
  {Polkovnikov}(2008)}]{Barankov_2008}%
  \BibitemOpen
  \bibfield  {author} {\bibinfo {author} {\bibfnamefont {R.}~\bibnamefont
  {Barankov}}\ and\ \bibinfo {author} {\bibfnamefont {A.}~\bibnamefont
  {Polkovnikov}},\ }\bibfield  {title} {\bibinfo {title} {{Optimal nonlinear
  passage through a quantum critical point}},\ }\href
  {https://doi.org/10.1103/PhysRevLett.101.076801} {\bibfield  {journal}
  {\bibinfo  {journal} {Phys. Rev. Lett.}\ }\textbf {\bibinfo {volume} {101}},\
  \bibinfo {pages} {076801} (\bibinfo {year} {2008})}\BibitemShut {NoStop}%
\bibitem [{\citenamefont {Gu{\'e}ry-Odelin}\ \emph {et~al.}(2019)\citenamefont
  {Gu{\'e}ry-Odelin}, \citenamefont {Ruschhaupt}, \citenamefont {Kiely},
  \citenamefont {Torrontegui}, \citenamefont {Mart{\'i}nez-Garaot},\ and\
  \citenamefont {Muga}}]{Muga_2019}%
  \BibitemOpen
  \bibfield  {author} {\bibinfo {author} {\bibfnamefont {D.}~\bibnamefont
  {Gu{\'e}ry-Odelin}}, \bibinfo {author} {\bibfnamefont {A.}~\bibnamefont
  {Ruschhaupt}}, \bibinfo {author} {\bibfnamefont {A.}~\bibnamefont {Kiely}},
  \bibinfo {author} {\bibfnamefont {E.}~\bibnamefont {Torrontegui}}, \bibinfo
  {author} {\bibfnamefont {S.}~\bibnamefont {Mart{\'i}nez-Garaot}},\ and\
  \bibinfo {author} {\bibfnamefont {J.~G.}\ \bibnamefont {Muga}},\ }\bibfield
  {title} {\bibinfo {title} {{Shortcuts to adiabaticity: Concepts, methods, and
  applications}},\ }\href {https://doi.org/10.1103/PhysRevLett.101.076801}
  {\bibfield  {journal} {\bibinfo  {journal} {Rev. Mod. Phys.}\ }\textbf
  {\bibinfo {volume} {91}},\ \bibinfo {pages} {045001} (\bibinfo {year}
  {2019})}\BibitemShut {NoStop}%
\bibitem [{\citenamefont {G.}\ and\ \citenamefont
  {Masanes}(2016)}]{Richens_2016}%
  \BibitemOpen
  \bibfield  {author} {\bibinfo {author} {\bibfnamefont {R.~J.}\ \bibnamefont
  {G.}}\ and\ \bibinfo {author} {\bibfnamefont {L.}~\bibnamefont {Masanes}},\
  }\bibfield  {title} {\bibinfo {title} {{Work extraction from quantum systems
  with bounded fluctuations in work}},\ }\href
  {https://doi.org/10.1038/ncomms13511} {\bibfield  {journal} {\bibinfo
  {journal} {Nat. Commun.}\ }\textbf {\bibinfo {volume} {7}},\ \bibinfo {pages}
  {13511} (\bibinfo {year} {2016})}\BibitemShut {NoStop}%
\bibitem [{\citenamefont {Holubec}\ and\ \citenamefont
  {Ryabov}(2017)}]{Holubec2017}%
  \BibitemOpen
  \bibfield  {author} {\bibinfo {author} {\bibfnamefont {V.}~\bibnamefont
  {Holubec}}\ and\ \bibinfo {author} {\bibfnamefont {A.}~\bibnamefont
  {Ryabov}},\ }\bibfield  {title} {\bibinfo {title} {{Work and power
  fluctuations in a critical heat engine}},\ }\href
  {https://doi.org/10.1103/PhysRevE.96.030102} {\bibfield  {journal} {\bibinfo
  {journal} {Phys. Rev. E}\ }\textbf {\bibinfo {volume} {96}},\ \bibinfo
  {pages} {030102(R)} (\bibinfo {year} {2017})}\BibitemShut {NoStop}%
\bibitem [{\citenamefont {Denzler}\ and\ \citenamefont
  {Lutz}(2020)}]{Denzler2020}%
  \BibitemOpen
  \bibfield  {author} {\bibinfo {author} {\bibfnamefont {T.}~\bibnamefont
  {Denzler}}\ and\ \bibinfo {author} {\bibfnamefont {E.}~\bibnamefont {Lutz}},\
  }\bibfield  {title} {\bibinfo {title} {{Efficiency fluctuations of a quantum
  heat engine}},\ }\href {https://doi.org/10.1103/PhysRevResearch.2.032062}
  {\bibfield  {journal} {\bibinfo  {journal} {Phys. Rev. Res.}\ }\textbf
  {\bibinfo {volume} {2}},\ \bibinfo {pages} {032062(R)} (\bibinfo {year}
  {2020})}\BibitemShut {NoStop}%
\bibitem [{\citenamefont {Denzler}\ and\ \citenamefont
  {Lutz}(2021)}]{Lutz2021}%
  \BibitemOpen
  \bibfield  {author} {\bibinfo {author} {\bibfnamefont {T.}~\bibnamefont
  {Denzler}}\ and\ \bibinfo {author} {\bibfnamefont {E.}~\bibnamefont {Lutz}},\
  }\bibfield  {title} {\bibinfo {title} {{Efficiency large deviation function
  of quantum heat engines}},\ }\href {https://doi.org/10.1088/1367-2630/ac09fe}
  {\bibfield  {journal} {\bibinfo  {journal} {New J. Phys.}\ }\textbf {\bibinfo
  {volume} {23}},\ \bibinfo {pages} {075003} (\bibinfo {year}
  {2021})}\BibitemShut {NoStop}%
\bibitem [{\citenamefont {Wiedmann}\ \emph {et~al.}(2020)\citenamefont
  {Wiedmann}, \citenamefont {Stockburger},\ and\ \citenamefont
  {Ankerhold}}]{Wiedmann_2020}%
  \BibitemOpen
  \bibfield  {author} {\bibinfo {author} {\bibfnamefont {M.}~\bibnamefont
  {Wiedmann}}, \bibinfo {author} {\bibfnamefont {J.~T.}\ \bibnamefont
  {Stockburger}},\ and\ \bibinfo {author} {\bibfnamefont {J.}~\bibnamefont
  {Ankerhold}},\ }\bibfield  {title} {\bibinfo {title} {Non-markovian dynamics
  of a quantum heat engine: out-of-equilibrium operation and thermal coupling
  control},\ }\href {https://doi.org/10.1088/1367-2630/ab725a} {\bibfield
  {journal} {\bibinfo  {journal} {New J. Phys.}\ }\textbf {\bibinfo {volume}
  {22}},\ \bibinfo {pages} {033007} (\bibinfo {year} {2020})}\BibitemShut
  {NoStop}%
\bibitem [{\citenamefont {Scully}\ \emph {et~al.}(2003)\citenamefont {Scully},
  \citenamefont {Zubairy}, \citenamefont {Agarwal},\ and\ \citenamefont
  {Walther}}]{Scully_2003}%
  \BibitemOpen
  \bibfield  {author} {\bibinfo {author} {\bibfnamefont {M.~O.}\ \bibnamefont
  {Scully}}, \bibinfo {author} {\bibfnamefont {M.~S.}\ \bibnamefont {Zubairy}},
  \bibinfo {author} {\bibfnamefont {G.~S.}\ \bibnamefont {Agarwal}},\ and\
  \bibinfo {author} {\bibfnamefont {H.}~\bibnamefont {Walther}},\ }\bibfield
  {title} {\bibinfo {title} {Extracting work from a single heat bath via
  vanishing quantum coherence},\ }\href
  {https://doi.org/10.1126/science.1078955} {\bibfield  {journal} {\bibinfo
  {journal} {Science}\ }\textbf {\bibinfo {volume} {299}},\ \bibinfo {pages}
  {862} (\bibinfo {year} {2003})}\BibitemShut {NoStop}%
\bibitem [{\citenamefont {Sau}\ and\ \citenamefont
  {Das~Sarma}(2012)}]{DasSarma_2012}%
  \BibitemOpen
  \bibfield  {author} {\bibinfo {author} {\bibfnamefont {J.~D.}\ \bibnamefont
  {Sau}}\ and\ \bibinfo {author} {\bibfnamefont {S.}~\bibnamefont
  {Das~Sarma}},\ }\bibfield  {title} {\bibinfo {title} {{Realizing a robust
  practical Majorana chain in a quantum-dot-superconductor linear array}},\
  }\href {https://doi.org/10.1038/ncomms1966} {\bibfield  {journal} {\bibinfo
  {journal} {Nat. Commun.}\ }\textbf {\bibinfo {volume} {3}},\ \bibinfo {pages}
  {964} (\bibinfo {year} {2012})}\BibitemShut {NoStop}%
\bibitem [{\citenamefont {Dvir}\ and\ \citenamefont {{\it et
  al.}}(2023)}]{Kouwenhoven_2023}%
  \BibitemOpen
  \bibfield  {author} {\bibinfo {author} {\bibfnamefont {T.}~\bibnamefont
  {Dvir}}\ and\ \bibinfo {author} {\bibnamefont {{\it et al.}}},\ }\bibfield
  {title} {\bibinfo {title} {{Realization of a minimal Kitaev chain in coupled
  quantum dots}},\ }\href {https://doi.org/10.1038/s41586-022-05585-1}
  {\bibfield  {journal} {\bibinfo  {journal} {Nature}\ }\textbf {\bibinfo
  {volume} {614}},\ \bibinfo {pages} {445} (\bibinfo {year}
  {2023})}\BibitemShut {NoStop}%
\bibitem [{\citenamefont {Breuer}\ and\ \citenamefont
  {Petruccione}(2007)}]{Petruccione_book}%
  \BibitemOpen
  \bibfield  {author} {\bibinfo {author} {\bibfnamefont {H.-P.}\ \bibnamefont
  {Breuer}}\ and\ \bibinfo {author} {\bibfnamefont {F.}~\bibnamefont
  {Petruccione}},\ }\href
  {https://doi.org/10.1093/acprof:oso/9780199213900.001.0001} {\emph {\bibinfo
  {title} {{The Theory of Open Quantum Systems}}}}\ (\bibinfo  {publisher}
  {Oxford University Press},\ \bibinfo {year} {2007})\BibitemShut {NoStop}%
\bibitem [{\citenamefont {Dorner}\ \emph {et~al.}(2012)\citenamefont {Dorner},
  \citenamefont {Goold}, \citenamefont {Cormick}, \citenamefont {Paternostro},\
  and\ \citenamefont {Vedral}}]{Dorner_2012}%
  \BibitemOpen
  \bibfield  {author} {\bibinfo {author} {\bibfnamefont {R.}~\bibnamefont
  {Dorner}}, \bibinfo {author} {\bibfnamefont {J.}~\bibnamefont {Goold}},
  \bibinfo {author} {\bibfnamefont {C.}~\bibnamefont {Cormick}}, \bibinfo
  {author} {\bibfnamefont {M.}~\bibnamefont {Paternostro}},\ and\ \bibinfo
  {author} {\bibfnamefont {V.}~\bibnamefont {Vedral}},\ }\bibfield  {title}
  {\bibinfo {title} {{Emergent thermodynamics in a quenched quantum many-body
  system}},\ }\href {https://doi.org/10.1103/PhysRevLett.109.160601} {\bibfield
   {journal} {\bibinfo  {journal} {Phys. Rev. Lett.}\ }\textbf {\bibinfo
  {volume} {109}},\ \bibinfo {pages} {160601} (\bibinfo {year}
  {2012})}\BibitemShut {NoStop}%
\bibitem [{\citenamefont {Fusco}\ \emph {et~al.}(2014)\citenamefont {Fusco},
  \citenamefont {Pigeon}, \citenamefont {Apollaro}, \citenamefont {Xuereb},
  \citenamefont {Mazzola}, \citenamefont {Campisi}, \citenamefont {Ferraro},
  \citenamefont {Paternostro},\ and\ \citenamefont {De~Chiara}}]{Fusco_2014}%
  \BibitemOpen
  \bibfield  {author} {\bibinfo {author} {\bibfnamefont {L.}~\bibnamefont
  {Fusco}}, \bibinfo {author} {\bibfnamefont {S.}~\bibnamefont {Pigeon}},
  \bibinfo {author} {\bibfnamefont {T.~J.}\ \bibnamefont {Apollaro}}, \bibinfo
  {author} {\bibfnamefont {A.}~\bibnamefont {Xuereb}}, \bibinfo {author}
  {\bibfnamefont {L.}~\bibnamefont {Mazzola}}, \bibinfo {author} {\bibfnamefont
  {M.}~\bibnamefont {Campisi}}, \bibinfo {author} {\bibfnamefont
  {A.}~\bibnamefont {Ferraro}}, \bibinfo {author} {\bibfnamefont
  {M.}~\bibnamefont {Paternostro}},\ and\ \bibinfo {author} {\bibfnamefont
  {G.}~\bibnamefont {De~Chiara}},\ }\bibfield  {title} {\bibinfo {title}
  {{Assessing the nonequilibrium thermodynamics in a quenched quantum many-body
  system via single projective measurements}},\ }\href
  {https://doi.org/10.1103/PhysRevX.4.031029} {\bibfield  {journal} {\bibinfo
  {journal} {Phys. Rev. X}\ }\textbf {\bibinfo {volume} {4}},\ \bibinfo {pages}
  {031029} (\bibinfo {year} {2014})}\BibitemShut {NoStop}%
\end{thebibliography}%

\end{document}